\documentclass[aps,pre,superscriptaddress,nofootinbib,preprintnumbers]{revtex4-2}

\usepackage{graphicx}
\usepackage{subfigure}
\usepackage{amsmath,amssymb}
\usepackage[usenames]{color}
\usepackage{hyperref}
\usepackage{breakurl}
\usepackage{enumitem}

\def\Tr{\,{\rm Tr}\,}

\renewcommand{\Re}[1]{\hbox{Re}~#1}

\newcommand{\be}{\begin{equation}}
\newcommand{\ee}{\end{equation}}
\newcommand{\bea}{\begin{eqnarray}}
\newcommand{\eea}{\end{eqnarray}}
\newcommand{\ben}{\begin{enumerate}}
\newcommand{\een}{\end{enumerate}}
\newcommand{\bit}{\begin{itemize}}
\newcommand{\eit}{\end{itemize}}

\newcommand{\la}[1]{\label{#1}}

\newcommand{\Eq}[1]{Eq.~(\ref{#1})}

\newcommand{\Sec}[1]{Sec.~\ref{#1}}

\newcommand{\Fig}[1]{Fig.~\ref{#1}}

\def\nl{\nonumber \\}

\newcommand{\vv}[1]{\mathbf #1}							
\newcommand\La{{\cal L}}
\newcommand\Na{{\cal N}}
\newcommand\Ua{{\cal U}}
\newcommand\Sa{{\cal S}}
\newcommand\Pa{{\cal P}}
\newcommand\Ja{{\cal J}}
\newcommand{\bert}{\raise-0.45mm\hbox{\Large$\Box$}}			

\newcommand{\gd}{\gamma_\downarrow}						
\newcommand{\gu}{\gamma_\uparrow}						

\definecolor{BrickRed}{cmyk}{0,0.89,0.94,0.28}					
\definecolor{MidnightBlue}{cmyk}{0.98,0.13,0,0.43}				
\definecolor{DarkGreen}{rgb}{0.100806,0.495968,0.209979}
\definecolor{orange}{rgb}{0.587167,0.354498,0.146197}

\begin{document}

\title{The problem of engines in statistical physics}
	
\author{Robert Alicki}
\email{robert.alicki@ug.edu.pl}
\affiliation{International Centre for Theory of Quantum Technologies (ICTQT), University of Gda\'nsk, 80-308, Gda\'nsk, Poland}
\author{David Gelbwaser-Klimovsky}
\email{dgelbi@mit.edu}
\affiliation{Physics of Living Systems, Department of Physics, Massachusetts Institute of Technology, Cambridge, MA 02139, USA}
\author{Alejandro Jenkins}
\email{alejandro.jenkins@ucr.ac.cr}
\affiliation{International Centre for Theory of Quantum Technologies (ICTQT), University of Gda\'nsk, 80-308, Gda\'nsk, Poland}
\affiliation{Laboratorio de F\'isica Te\'orica y Computacional, Escuela de F\'isica, Universidad de Costa Rica, 11501-2060, San Jos\'e, Costa Rica}


\begin{abstract}

Engines are open systems that can generate work cyclically, at the expense of an external disequilibrium.  They are ubiquitous in nature and technology, but the course of mathematical physics over the last 300 years has tended to make their dynamics in time a theoretical blind spot.  This has hampered the usefulness of statistical mechanics applied to active systems, including living matter.  We argue that recent advances in the theory of open quantum systems, coupled with renewed interest in understanding how active forces result from positive feedback between different macroscopic degrees of freedom in the presence of dissipation, point to a more realistic description of autonomous engines.  We propose a general conceptualization of an engine that helps clarify the distinction between its heat and work outputs.  Based on this, we show how the external loading force and the thermal noise may be incorporated into the relevant equations of motion.  This modifies the usual Fokker-Planck and Langevin equations, offering a thermodynamically complete formulation of the irreversible dynamics of simple oscillating and rotating engines. \\


\end{abstract}

\maketitle

\tableofcontents

\newpage


\section{Introduction}
\la{sec:intro}

An {\it engine} is as an open system that can undergo a cyclical transformation (i.e., one in which the macroscopic initial and final states are the same) that produces net work $W > 0$.  The energy necessary to perform this work comes from an external disequilibrium.  When that external disequilibrium is thermal (i.e., a difference of temperatures), we call the open system a {\it heat engine}.  This is consistent with the definition of heat engine used by all modern thermodynamic textbooks (see, e.g., \cite{Schroeder}).  When the external disequilibrium is not thermal we will refer to the open system as a {\it chemical engine}, since the disequilibrium can be interpreted as a difference of chemical potentials.

Engines are ubiquitous in nature and technology.  Life in general and our own human way of life in modern society depend on the ability of engines to do work cyclically, retaining their macroscopic structure as matter and/or heat passes through their boundaries.  Nonetheless, the {\it dynamics in time} of engines has long constituted a blind spot in theoretical physics.  Among theorists, autonomous engines have, for the most part, been treated in ways that explicitly avoid consideration of the time-dependence of the engine's macroscopic state or else described using physically unrealistic concepts such as negative friction, retarded action, resonance under an external periodic driving, or even Maxwell demons.

These conceptual difficulties have deep roots in the intellectual course of mathematical physics during the past three centuries.  The analytical mechanics of Euler, Lagrange, Hamilton, and Jacobi is only applicable to closed (i.e., conservative) systems or, with some modifications, to mechanical systems that exhibit simple damping.  The phenomenological thermodynamics developed in the 19th century, based on the work of Carnot, Joule, Helmholtz, Clausius, Kelvin, Maxwell, and Gibbs, was to a large extent a science of engines, but one in which time is never considered as a variable.  When Maxwell, Boltzmann, and Gibbs explained the laws of thermodynamics in statistical terms in the late 19th and early 20th centuries, they provided us with mathematical formalisms that could be applied to systems in macroscopic equilibrium, fluctuating stochastically about equilibrium, or passively relaxing towards equilibrium, but not to autonomous engines.

 It is only in recent years and within the specific context of quantum thermodynamics that the dynamics in time of engines has begun to attract serious attention in theoretical physics and chemistry; see \cite{QT} and references therein.  There is also an independent line of research focused on current-induced forces (see \cite{current} and references therein) and their application to adiabatic quantum motors \cite{AQM1, AQM2}, which has addressed a similar question from a somewhat different theoretical perspective.  Unfortunately, there has been little dialogue so far between these approaches.  Moreover, both of these lines of investigation have remained focused on mesoscopic quantum engines, with little consideration given to their relation with classical macroscopic engines.
 
An engine's work output $W$ is performed as the {\it working substance} within the engine exerts an {\it active force} upon a macroscopic degree of freedom, which here we will generically call the {\it tool}.  The cyclical nature of the engine's operation ensures that this tool does not ``run away'', either in position or in some more generalized macroscopic configuration space.  Though more complicated arrangements are possible in principle, here we will only be concerned with tools that can be characterized as either an oscillating {\it piston} or a rotating {\it turbine}.  Thus, throughout this article the word {\it tool} can be substituted by ``piston or turbine''.

The {\it load} is the system outside of the engine upon which the tool acts.  This load absorbs the engine's work output, preventing unbounded accumulation of mechanical energy in the tool while the engine is running.  We will argue that the distinction between the internal tool and the external load helps to distinguish clearly between the heat and work outputs of an autonomous engine.  When the engine runs steadily, its work output corresponds to the zero-entropy energy that the load extracts from the tool.  This is distinct from the dissipation into heat of the tool's mechanical energy by the damping and which is connected to thermal noise by a fluctuation-dissipation relation \cite{Kubo}.

Engines (or, in some contexts, only their corresponding loads) may be described as {\it active systems}, by contrast to closed systems subject only to conservative forces, and also to {\it passive systems} in which the nonconservative forces are either dissipative or stochastic and cannot do net positive work.  The terms {\it motor} and {\it prime mover} may be regarded as synonymous with {\it engine}, at least for the purposes of the present discussion.  The use of the term {\it machine} is more problematic, as it is also commonly applied to passive devices such as levers and pulleys.  An engine is necessarily an active system, but we will reserve the term ``engine'' for relatively simple systems that follow regular and mostly deterministic cycles.  Active systems in general may be regular, chaotic, or intermittent, as is evident from biophysics.

The fundamental theorem of calculus implies that a force proportional to the gradient of any generalized potential must do zero net work over a course of a cyclical transformation.  The active force that drives the engine's tool cannot, therefore, be obtained from any time-independent potential.  This applies not only to mechanical potentials, but also to thermodynamic state functions.  Onsager's formulation of non-equilibrium thermodynamics, including active systems like thermoelectric generators \cite{Onsager1, Onsager2}, is therefore not applicable to a dynamical description of autonomous engines.

There have been extensive and long-standing efforts to formulate non-equilibrium thermodynamics in terms of the extremization by the open system of the rate of entropy production, given a set of external constraints; see \cite{MEPP} and references therein.  Such maximum or minimum entropy production principles may apply to passive or externally driven systems, but not to engines.  The reason is that, for an irreversible cyclical transformation, the entropy production rate cannot be represented by fluxes or forces that depend only on local thermodynamic state variables.  Moreover, the stability of the cycle cannot be determined by an equality of free energies.  Already in 1975, Landauer showed the inapplicability to ``active processes'' in electrical circuits of Prigogine et al.'s theorems on the stability of dissipative structures \cite{Landauer1}; see also Landauer's argument against the generality of such extremal entropy-production principles in \cite{Landauer2}.  Landauer saw clearly that these principles fail when the system cannot be wholly characterized in terms of potentials because there is a {\it kinetic energy} (associated with a form of inertia, which is not necessarily mechanical) that cannot be neglected.  In our treatment of engines, this is the tool's kinetic energy.

In modern non-equilibrium thermodynamics, Prigogine and his collaborators conceptualized some active systems as ``dissipative structures'' \cite{Prigogine}.  At around the same time, Haken proposed an independent but conceptually similar ``synergetic'' treatment, which was largely inspired by the active dynamics of lasers \cite{Haken}.  That work met with some strong objections (see, e.g., \cite{Anderson}) that, in our view, were justified by the inability of such formalisms to describe the dynamics in time of such structures operating as autonomous engines \cite{LEC}.

Our own perspective is based on the principle that the work output of an engine should be understood in terms of the active force that the working substance exerts on the macroscopic tool (piston or turbine).  This active force results from a {\it positive feedback} between the mechanical state of the tool and the coupling of the working substance to the external disequilibrium, from which energy can be extracted in a thermodynamically {\it irreversible} way.  The active force is therefore a function not just of the working substance's thermodynamic state, but also of the mechanical degrees of freedom of the tool on which the active force does net positive work over a full thermodynamic cycle.  As we will see, this formulation leads to homogenous (i.e., autonomous) equations of motion in which the active force is not given by the gradient of {\it any} potential function.

Open systems that generate and maintain the periodic motion of a piston, at the expense of a source of power that has no corresponding periodicity, have been labelled {\it self-oscillators} \cite{AVK}.\footnote{The same concept is identified elsewhere in the literature as ``maintained'', ``sustained'', ``autonomous'', ``self-maintained,'' ``self-sustained,'' ``self-sustaining,'' or ``self-exciting'' oscillators.  The prefix ``self'' may be substituted by ``auto''.}  There is a rich literature on the subject that stretches back to the early work of Airy on the operation of the vocal cords \cite{Airy} and of Rayleigh on non-percussive musical instruments \cite{ToS}, but this literature has been largely mathematical and based on the theory of ordinary differential equations and control theory (see, e.g., \cite{AVK, Strogatz, Kirillov}).  In the 1930s and '40s, Le Corbeiller stressed that self-oscillators are engines and advocated connecting their mathematical theory with thermodynamics \cite{LeC1, LeC2}, but he did not pursue this line of investigation very far and it had little impact in the subsequent scientific literature.\footnote{Le Corbeiller used the the term ``motor'' in \cite{LeC1} and ``prime mover'' in \cite{LeC2}.  Our use of the word ``tool'' in the context of engine dynamics is borrowed directly from \cite{LeC2}.}  Much later, that approach was advocated and explored in \cite{SO}.

Little theoretical work has been done on {\it self-rotors} \cite{Kirillov, Lugt}, in which the active force turns a turbine.  Self-rotors appear rather different from self-oscillators when considered mathematically, but are similar from the point of view of the thermodynamics of engines \cite{rotors}.  Even in the seemingly trivial problem of the operation of a waterwheel, turned by a flow powered by gravity, dissipation and feedback are needed to understand the active force that turns the wheel \cite{current}.  Windmills are turned by aerodynamic lift \cite{Bob-lift}, an active force that depends on the air's viscosity and for which the details of the feedback are still debated by experts \cite{SA-lift}.  In \Sec{sec:Quincke} we will work out in detail the active dynamics of a simple rotating engine (the ``Quincke rotor'') powered by a steady electrical current.

The problem of engines in statistical physics is further complicated by the fact that the tool is hidden in many active systems of interest, both in engineering (e.g., in batteries, fuel cells, thermoelectric generators, and photovoltaic cells) and in biology (e.g., in ion pumps and in excitable cell membranes).  We believe that this is because those systems have an electrical double layer (EDL) that acts as a self-oscillating piston, moving with such a high frequency and small amplitude that the oscillation has thus far generally evaded direct detection.  In such cases the smallness of the amplitude of the oscillation may be compensated by the large surface area of the EDL.  When experimental evidence of such an oscillation has been reported (e.g., in \cite{Nakanishi, Guzelturk}), it has not usually been interpreted as an essential part of the engine dynamics of the corresponding active systems, and the feedback mechanism that gives rise to the corresponding active force has not been investigated.  On this issue, see \cite{LEC, solarcells, thermocells, fuelcells, AGJ, Torun, solar-dyn, battery}, by the authors and their collaborators.  Independently, Goupil et al.\ have argued for the importance of feedback-induced periodicity in describing the operation of thermoelectric generators \cite{Goupil}.

Finally, another outstanding problem in the theory of engines is how to incorporate thermal fluctuations into their dynamics.  Such fluctuations are not relevant for macroscopic engines with heavy tools, but as the mass of the tool is reduced the fluctuations become more important, and there must be a limit beyond which the feedback dynamics responsible for the active force is swamped by the thermal noise, so that the engine can no longer operate.  This has recently been investigated in the context of stochastic thermodynamics \cite{stochastic-shuttle, shuttle-rotor, Strasberg}.  Our treatment will be somewhat different and based on drawing a distinction between damping forces associated with thermal noise and the force that the load exerts on the tool (which is not associated with thermal fluctuations).  This will allow us to write the equations of motion for some simple oscillating and rotating engines in terms of modifications of the Fokker-Planck and Langevin equations as they have been applied previously to passive or externally driven systems in contact with thermal baths.

The plan for this article is the following:  Secs.\ \ref{sec:feedback} and \ref{sec:fluctuations} are largely reviews of previously published results, but organized and presented in a novel way that seeks to integrate that material into a general dynamical treatment of engines that connects naturally with statistical physics.  The focus here is on justifying and clarifying our approach, which is qualitatively different from the formulations of non-equilibrium thermodynamics that have been predominant in the literature since the work of Onsager in the 1930s \cite{Onsager1, Onsager2}.  The treatment of electrostatic engines in \Sec{sec:electrostatic} is partly original and intended to provide simple applications of our approach that can also be of interest in biophysics and nanotechnology.  Section \ref{sec:dynamics} offers a new formulation of engine dynamics, based on the master equation for open systems, allowing us to characterize the engine's steady operation under a load and to distinguish clearly between its heat and work outputs while incorporating thermal noise.  Here and in the final discussion of \Sec{sec:discussion}, we also formulate and stress some of the most important open questions in a field that is only beginning to emerge as a distinct research program in theoretical physics and chemistry.


\section{Feedback in classical engines}
\la{sec:feedback}

In thermodynamics textbooks, the equilibrium values of the various thermodynamic variables are related to each other mathematically, but these are never expressed as functions of time.  The operation of engines is described in terms of cycles, but attention is focused on the reversible limit (e.g., the Carnot cycle) in which the transformations of the working substance are infinitely slow \cite{Schroeder}.  Recent work on ``finite-time thermodynamics'' has stressed that an engine capable of generating positive power must operate irreversibly; see \cite{Ouerdane} and references therein.  However, it is rare to find a treatment of the operation of an autonomous heat engines in terms of any {\it equation of motion} (i.e., a differential equation with time as the independent variable and the mechanical degrees of freedom as the dependent variables).  For this reason, the nature of the {\it active force} that drives the engine's tool is rarely considered.  When it is taken into account, it is usually not in a physically realistic way.

Active forces result from positive feedback between distinct macroscopic degrees of freedom, one of which acts as the engine's tool.  Such feedback can exist only in non-conservative systems \cite{dissipation-induced}.  The reversible operation of an engine (as in the idealized Carnot cycle) corresponds to a limit in which this active force vanishes.  As we will see, the need to consider feedback in a dynamical description of autonomous engines can be understood both thermodynamically and mechanically.


\subsection{Rayleigh-Eddington criterion}
\la{sec:RE}

The need to consider the {\it dynamics in time} of an engine in order to describe it in a physically realistic way can be deduced directly from the first and second laws of thermodynamics, even if they do not directly allow us to arrive at equations of motions.  To see this, consider an infinitesimal transformation of the engine's homogenous working substance, for which $\delta W$ is the work done upon the tool, $\delta Q$ is the heat absorbed from the surroundings, and $d {\cal N}$ the change in the quantity of matter.\footnote{The symbol $d$ is used for exact differentials, while $\delta$ is used for inexact differentials.  Throughout this article, we use calligraphic letters for extensive thermodynamic state variables.}  By the first law of thermodynamics, the change in the internal energy $\cal U$ is
\be
d{\cal U} = \delta Q - \delta W + \mu \, d {\cal N} ,
\la{eq:1stlaw}
\ee
where $\mu$ is the chemical potential.  Some authors refer to $\mu d {\cal N}$ as ``chemical work'', but the distinction between the actual {\it mechanical} work $\delta W$ (associated with a force acting on a {\it directional} degree of freedom) and this chemical contribution to the internal energy is one that we wish to maintain and stress throughout.  Although the chemical contribution carries no entropy, its {\it cyclical} conversion into mechanical work is non-trivial and will require dissipation and feedback, leading to an active force capable of driving a tool; see, e.g., \cite{current, LeC1, LeC2, Jaynes}.

Over a complete thermodynamic cycle the substance returns to its initial state, so that the change to the internal energy $\cal U$ vanishes:
\be
\Delta {\cal U} = \oint \left( \delta Q - \delta W + \mu \, d {\cal N} \right) = 0 .
\la{eq:conservation}
\ee
The net work is therefore
\be
W = \oint \delta W = \oint ( \delta Q + \mu \, d {\cal N}) .
\la{eq:work}
\ee
By Clausius's theorem (which is a formulation of the second law of thermodynamics) the cycle produces entropy
\be
\Sigma = - \oint \frac{\delta Q}{T} = - \oint \frac{\delta Q}{\bar T \left ( 1 + T_d / \bar T \right)} \geq 0 ~,
\la{eq:Clausius}
\ee
where $T$ is the substance's instantaneous temperature, with $\bar T$ its mean value over the cycle's period and \hbox{$T_d \equiv T - \bar T$}.  Let $\mu$ be the instantaneous chemical potential and $\bar \mu$ its mean, with \hbox{$\mu_d \equiv \mu - \bar \mu$}.  Adding \Eq{eq:Clausius} to \Eq{eq:work}, we obtain
\be
W \leq \oint \delta Q \left( 1 - \frac{1}{1 + T_d / \bar T} \right) + \oint d {\cal N} \cdot \mu = \oint \frac{\delta Q \cdot T_d}{\bar T + T_d} + \oint d {\cal N} \cdot \mu_d ~.
\la{eq:Rayleigh}
\ee
This bound on $W$ is achieved by a reversible cycle ($\Sigma = 0$).  The result of \Eq{eq:Rayleigh} is easily generalized to inhomogenous temperatures and chemical potentials by integrating over the maximum work that each part of the working substance may perform.

For a pure heat engine ($\mu_d = 0$), \Eq{eq:Rayleigh} simplifies to
\be
W \leq \oint \frac{\delta Q \cdot T_d}{\bar T + T_d} \simeq \frac{1}{\bar T} \oint \delta Q \cdot T_d ~,
\la{eq:Eddington}
\ee
where in the last step we assumed that $| T_d | \ll \bar T$ throughout the cycle (as is generally the case for heat engines that operate far from absolute zero). Eddington wrote \Eq{eq:Eddington} for his theory of the self-oscillation of Cepheid variable stars \cite{Eddington1, Eddington2}.  Before him, Rayleigh had formulated a similar criterion for thermoacoustic phenomena: that the oscillation of an element of gas is most encouraged when the rate of heat flowing out of it varies in phase with its volume \cite{Rayleigh1, Rayleigh2}.  For a substance with positive pressure and heat capacity, adiabatic expansion reduces its temperature and adiabatic compression increases it, so that Rayleigh's criterion follows from \Eq{eq:Eddington}.  Even earlier, Beau de Rochas had made a breakthrough in the design of internal combustion engines by proposing (based on a mechanical rather than thermodynamic argument) that the fuel's ignition be timed to coincide precisely with the maximum compression of the working gas in the cylinder \cite{Rochas}.  Modern diesel and gasoline engines work on this principle \cite{motors}.

\begin{figure} [t]
\begin{center}
	\includegraphics[width=0.4 \textwidth]{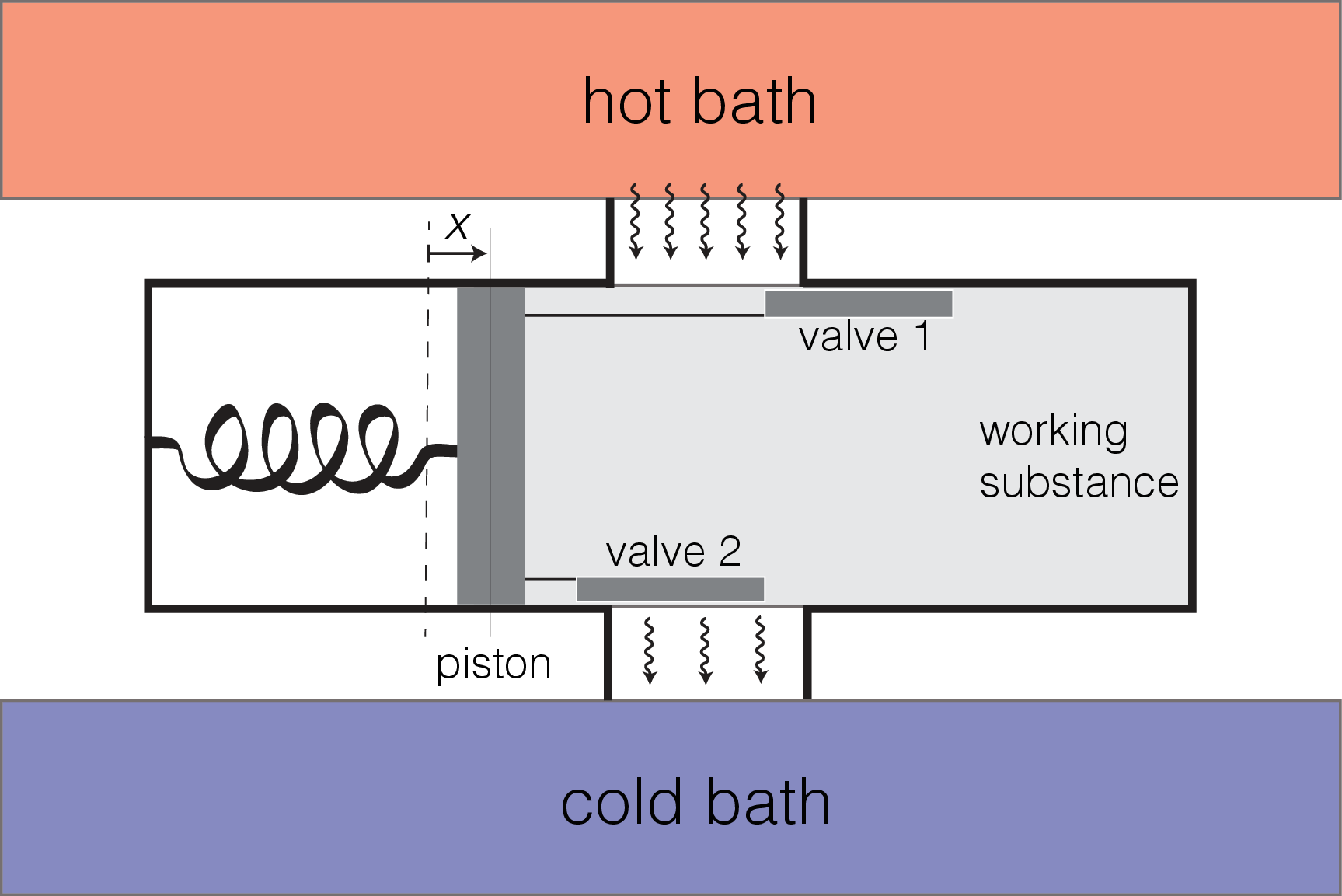}
\end{center}
\caption{\small The autonomous operation of this heat engine depends on the valves modulating the rate of heat flow between the working substance and the two baths in accordance with \Eq{eq:Eddington}, so that $W > 0$.  This gives a positive feedback between the oscillation of $x$ and the modulation of the heat currents.  Image taken from \cite{AGJ}.\la{fig:Rayleigh}}
\end{figure}

Let $x$ be the displacement of the piston, with increasing $x$ corresponding to compression of the working substance, as shown in \Fig{fig:Rayleigh}.  For the engine to run ($W > 0$), the relative phase $\varphi$ between $\delta Q$ and $x$ (or, equivalently, between $\delta Q$ and $T_d$ in \Eq{eq:Eddington}) must satisfy $-90^\circ < \varphi < 90^\circ$.  We call this the ``Rayleigh-Eddington criterion'' \cite{AGJ}.  When heat flow is modulated by the piston's motion in accordance with this criterion, a {\it positive feedback} is established, which, if it overcomes the passive damping, causes the piston to self-oscillate.  Heat is most efficiently converted into mechanical energy when $\varphi = 0$.  In the engine in \Fig{fig:Rayleigh}, this is achieved by arranging for valve 1 to be open and valve 2 closed when $x$ reaches its maximum.  Conversely, valve 1 should be closed and valve 2 open when $x$ reaches its minimum.

Note that, by reversing the signs of $\delta Q$ and $W$ in \Eq{eq:Eddington}, we may deduce that a cycle that violates the Rayleigh-Eddington criterion can run as a refrigerator, extracting heat from the working substance at the expense of external mechanical power.  The sign of the frequency of the piston's oscillation in \Fig{fig:Rayleigh} is not physically meaningful, and running it as a refrigerator would require interchanging the configurations of valves 1 and 2.  In some implementations of a heat cycle, such as Stirling engines, reversal between engine and refrigerator can be accomplished by reversing the direction in which the flywheel rotates \cite{Stirling}.  In that case the sign of the flywheel's angular velocity is given physical significance by the phase relation between the motion of the power piston and the rate of heat injection to the working substance.  The mathematical distinction between oscillating and rotating engines will recur in various forms throughout the rest of our discussion.

According to \Eq{eq:Rayleigh}, a cycle of an engine that operates between baths at different values of $\mu$ may do positive work at constant temperature ($T_d = 0$), as long as the working substance's $\mu$ is modulated so that
\be
0 < W \leq \oint d {\cal N} \cdot \mu_d .
\la{eq:mu-modulation}
\ee
That is, a chemical engine must take in matter at higher chemical potential and expel it at lower chemical potential.  We refer to \Eq{eq:mu-modulation} as the ``chemical Rayleigh-Eddington criterion'', although neither Rayleigh nor Eddington applied their reasoning to chemical engines.  This simple criterion is useful in understanding the dynamics of the engines that we consider \Sec{sec:electrostatic}.

Since an engine is an open system, the analysis of its dynamics will depend on how we choose to draw its boundary.  As was pointed out in \cite{battery}, an internal combustion engine can be conceptualized either as a heat engine, with the air in the cylinder as the working substance, or as a chemical engine, with the fuel-air mixture as the working substance.  In the latter analysis, \Eq{eq:mu-modulation} tells us that the engine works because it takes in pristine fuel ($\mu_d > 0$) and expels burnt fuel ($\mu_d < 0$).  Note that \Eq{eq:mu-modulation} gives a valid but loose upper bound on a combustion engine's work output, because of the heat that the engine dumps into its environment when it expels the burnt fuel, making the entropy output $\Sigma$ large for the corresponding cycle.

As we already noted in \Sec{sec:intro}, the engine's entropy production rate $\dot \Sigma$ cannot be expressed in terms of fluxes and forces depending only on local thermodynamic state variables, and it therefore does not fit into the usual Onsagerian picture of irreversible processes \cite{Onsager1, Onsager2, MEPP}.  This is because the tool (such as the piston in \Fig{fig:Rayleigh}) is a macroscopic degree of freedom endowed with inertia (and therefore with kinetic energy) whose motion affects, via a feedback, the coupling of the working substance to the external baths.  This means an engine's dynamics cannot be entirely characterized in terms of thermodynamic potentials, also rendering inapplicable the maximum or minimum entropy production principles that can be used to describe passively irreversible processes \cite{Landauer1, Landauer2}.  Instead, we will seek to understand the engine's active dynamics using irreversible equations of motion that explicitly incorporate the relevant feedback and active force.


\subsection{Active force}
\la{sec:active}

The Rayleigh-Eddington criterion of \Sec{sec:RE} establishes that an engine can do net positive work without ``running away'' in thermodynamic state space only if the state of the working substance varies cyclically so that the substance is at a higher temperature when it absorbs heat and at a lower temperature when it rejects heat,  and/or if matter is added to the working substance at higher chemical potential and removed at lower chemical potential.  If the engine runs autonomously, there must be a feedback mechanism that regulates the flow of heat and/or matter in phase the temperature and/or chemical potential of the working substance.

In practice, such a feedback always results from the way in which the state of the working substance depends on the configuration of the same tool upon which that substance exerts the {\it active force}.  Since thermodynamic textbooks are not usually concerned with the dynamics in time of the engine's cycle, the nature of this feedback is often overlooked in theoretical treatments of engines.  When active forces have been introduced in the literature, it has usually been in effective terms that cannot be applied realistically to an autonomous engine.  Such terms include a negative friction or anti-damping (as in the Rayleigh \cite{Rayleigh-vdP}, van der Pol \cite{vdP1, vdP2}, and Stuart-Landau \cite{Stuart-Landau} models, as well as in models of self-propelled Brownian particles \cite{active-brownian, self-propelled}), a circulatory force (see, e.g., the treatment of the nonconservative force in stochastic thermodynamics \cite{Seifert}), a time delay in an oscillator's response \cite{Airy, Minorsky, bio-oscillators}, or as an external periodic forcing (see, e.g., the model of Brownian motors in \cite{Hanggi}).  In \Sec{sec:deterministic} we will see how these can be regarded as effective descriptions of the dynamical effects of a physical feedback.

A rare instance of a realistic description of an autonomous heat engine in terms of classical equations of motion, with the active force arising from a positive feedback between the state of the tool and that of the working substance, is found in the pedagogical treatment of the putt-putt steamboat by Finnie and Curl \cite{pp1, pp2}.\footnote{The ``putt-putt'' or ``pop-pop'' steamboat was once a popular children's toy, and remains so in India and some other parts of the world.  See Sec. 7.2.1 in \cite{SO} for further details.}  Let us briefly review this analysis, using the variation of the putt-putt engine introduced in \cite{AGJ} and illustrated in \Fig{fig:putt-pump}.   This will serve us as a useful paradigm for studying the dynamics of other engines.

\begin{figure} [t]
\begin{center}
	\includegraphics[width=0.42 \textwidth]{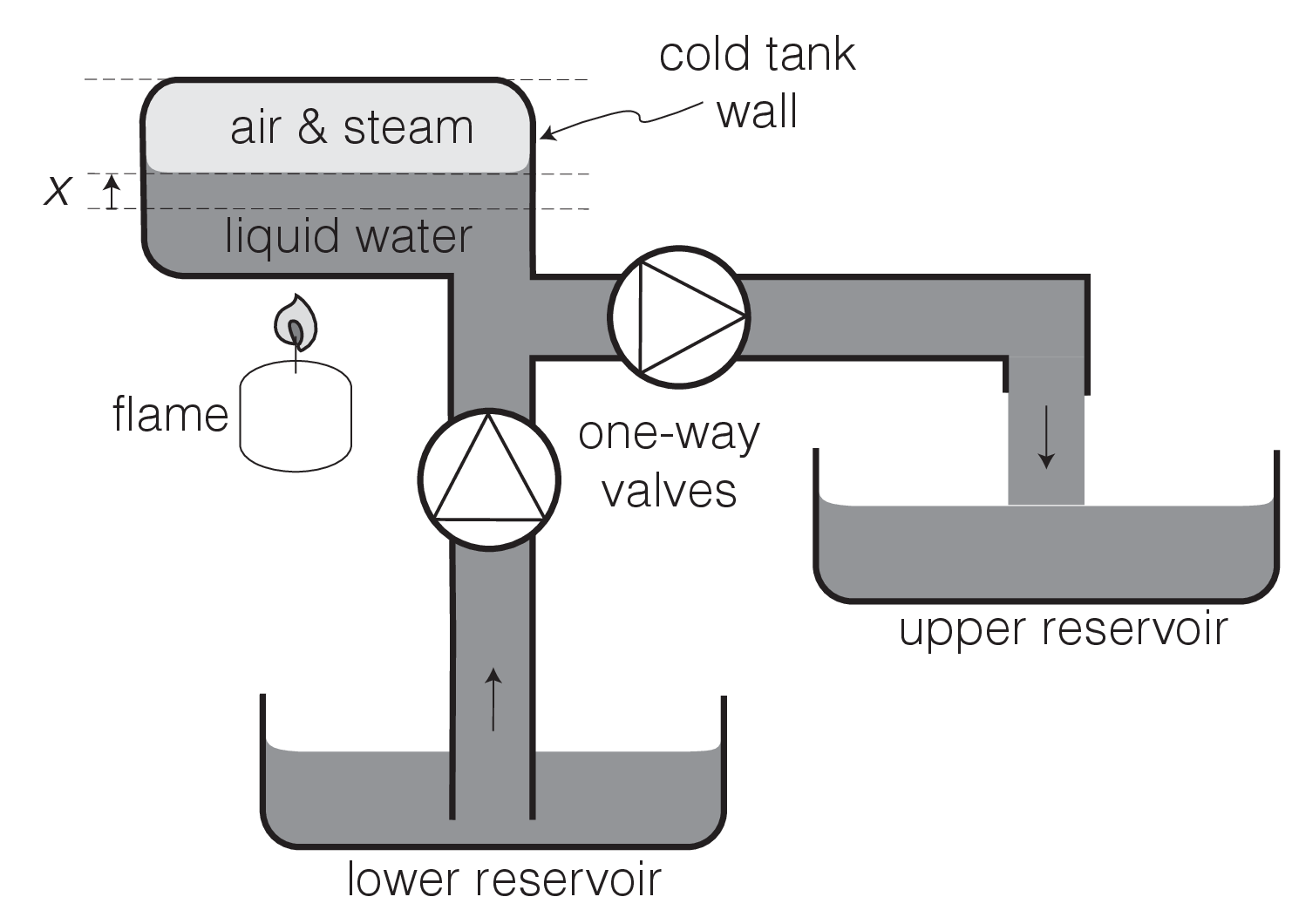}
\end{center}
\caption{\small Model of an autonomous ``putt-putt'' heat engine that can raise water from a lower to an upper reservoir.  Image taken from \cite{AGJ}.\la{fig:putt-pump}}
\end{figure}

For the putt-putt engine, the working substance is the bubble of air and steam in the tank (see \Fig{fig:putt-pump}).  This gas acts upon the surface of the liquid water, whose height $x$ we measure relative to the equilibrium position.  Thus, $x$ can be regarded as the tool degree of freedom.  The load is represented in \Fig{fig:putt-pump} by the pumping of water from the lower to the upper reservoirs, which builds up a gravitational potential.

Let $N_0$ be the amount of steam at equilibrium.  Most of the steam remains at the same temperature as the liquid water below it.  Small oscillations of $x$ therefore obey
\be
\ddot x + \gamma \dot x + \omega^2 x =  A \left( N_0 - N \right) ,
\la{eq:x} 
\ee
where $A$ is positive and constant, and $N$ is the instantaneous quantity of steam.  The rate of change in the quantity of steam may be expressed as
\be
\dot N  =  - \Gamma(x) {N} + B(x) ,
\la{eq:N}
\ee
where $\Gamma (x) \geq 0$ is the condensation rate and $B(x) \geq 0$ is the evaporation rate.  For appropriate values of the parameters of this model, the heat from the flame can sustain the self-oscillation of $x$ about its equilibrium.

To find the conditions under which $x$ will self-oscillate, we linearize Eqs.\ \eqref{eq:x} and \eqref{eq:N}, putting
\be
\Gamma(x) = \Gamma  + g x , \quad B(x) = B + b x , \quad \hbox{and} \quad n \equiv N - N_0 ~,
\la{eq:lin}
\ee
with $N_0 = B / \Gamma$.  In units of time and position such that $\omega =1 $ and $A = 1$, the linearization of Eqs.\ \eqref{eq:x} and \eqref{eq:N} can be expressed as
\be
\left( \begin{array}{c} \dot x  \\ \dot v \\ \dot n \end{array} \right) = 
\left( \begin{array}{r r r}
0 & 1 & 0 \\
-1 & - \gamma & -1 \\
f & 0 & -\Gamma 
\end{array} \right)
\left( \begin{array}{c} x \\ v \\ n \end{array} \right) ,
\la{eq:matrix}
\ee
with feedback parameter
\be
f \equiv b - g N_0 .
\la{eq:b}
\ee
The characteristic equation for the $3 \times 3$ matrix in \Eq{eq:matrix} is
\be
[ \lambda (\lambda + \gamma) +1 ] (\lambda + \Gamma) + f = 0 ~.
\la{eq:poly}
\ee
When the real part of an eigenvalue $\lambda$ is positive, the equilibrium $x=0$ is unstable and the amplitude of small perturbations grows exponentially with time.  If that eigenvalue has non-zero imaginary part, this corresponds to a self-oscillation.  In the mathematical literature, the transition from $\Re \lambda < 0$ to $\Re \lambda > 0$ is often called a ``Hopf bifurcation'' \cite{Strogatz}.

Approximate solutions of \Eq{eq:poly} can be found analytically for $\Gamma, \gamma , |f| \ll 1$, in which case there will be one real eigenvalue $\lambda_0$ and two complex, conjugate eigenvalues $\lambda_{\pm}$ close to $\pm i$. Then, up to higher-order corrections,
\be
\lambda_0 \simeq - \left( \Gamma + f \right) \quad \hbox{and} \quad
\lambda_{\pm} \simeq \pm i +\frac{1}{2} \left( f - \gamma \right) ~.
\la{eq:eigen}
\ee
Self-oscillation therefore occurs for a positive feedback parameter $f > \gamma$.  In physical units, this corresponds to
\be
f = \frac{A \left( b - g N_0 \right)}{\omega^2} >  \gamma .	
\la{eq:self_cond1}
\ee
The damping coefficient $\gamma$ in \Eq{eq:x} must be positive, since it corresponds to the irreversible dissipation of mechanical energy into heat.  For $\Gamma, \gamma,$ and $|f|$ not small, the eigenvalues may be computed numerically, as discussed in \cite{AGJ}.\footnote{See also the stability analysis of the ``leaking elastic capacitor'' model in \cite{LEC}.}

This analysis of the putt-putt heat engine, based on deterministic equations of motion for macroscopic degrees of freedom, may appear to be wholly mechanical.  But the presence of non-conservative forces in the equations of motion requires a proper thermodynamic justification.  According to Landau and Lifshitz, ``the problem of the motion of a body in a medium is not one of mechanics'', since the non-conservation must be accounted for thermodynamically \cite{LL}.\footnote{On the microphysical interpretation of the pressure that the steam molecules exert on the liquid surface (which corresponds to the right-hand side of \Eq{eq:x}) and its connection with surface tension, see \cite{AGJ}.}

In the putt-putt engine, latent heat is injected into the working substance by the evaporation of the liquid water, while latent heat is rejected by the condensation of water vapor.  The condition $f > 0$ in \Eq{eq:self_cond1} can be translated to the physical requirement that net evaporation should increase as $x$ increases, while net condensation should increase as $x$ decreases.  According to Finnie and Curl, in practice the evaporation rate is probably nearly independent of $x$, but the condensation rate increases significantly when $x$ is negative enough that a significant surface area of the tank's metal wall (which has been cooled by the liquid water that is further from the flame) is exposed to the working gas \cite{pp1, pp2}.  This is consistent with the formulation of the Rayleigh criterion in terms of the volume of the working gas: the heat outflow (i.e., condensation) is modulated in phase with the total volume of the working gas.  We conclude that the dynamical analysis presented above is consistent with the Rayleigh-Eddington criterion of \Sec{sec:RE}, and therefore with the laws of thermodynamics.

In this case we can see how the active force arises from a positive feedback between $x$ and $n$.  When the condition of \Eq{eq:self_cond1} is met, this active force causes self-oscillation of $x$.  In the linear regime the amplitude of the self-oscillation grows exponentially without bound.  We must therefore either introduce nonlinear damping terms\footnote{The thermodynamic and statistical interpretation of such nonlinear damping is obscure, as we point out in \Sec{sec:discussion}.} and/or take into account the extraction of work by the load (represented by the accumulation of water in the upper reservoir in \Fig{fig:putt-pump}) in order to explain how the self-oscillation approaches a {\it limit cycle} with finite amplitude \cite{SO}.

Since water is lifted only during the $\dot x > 0$ phase of the self-oscillation, the work extraction by the load must be represented in the equation of motion (\Eq{eq:x}) as a velocity-dependent force, which will therefore be non-conservative.  The problem lies in the dynamics of the one-way valves in \Fig{fig:putt-pump}, which we will not attempt to work out here.  The need for {\it rectification} introduces complications in the modeling of the loading force for oscillating engines that we will revisit in \Sec{sec:stochastic-osc}.  In this respect, it will be easier to describe the loading of a rotating engine, which can be easily represented as a constant (and therefore conservative) torque.


\subsection{Electrostatic engines}
\la{sec:electrostatic}

The first electrostatic engines were built in the early 1740s by a Scottish Benedictine monk, Andrew Gordon, who was a professor of philosophy as the University of Erfurt.  Gordon built both a self-oscillating engine (``electric bells'') and a self-rotating one (``electric whirl'').  Both of these were soon modified and improved by Benjamin Franklin, who is therefore often credited with their development.  For a history of these devices, see \cite{Jefimenko}.   

The source of energy in such electrostatic engines is a fixed external voltage difference $V_0$, which can be represented thermodynamically as a difference in the chemical potential $\Delta \mu = e V_0$, for elementary charge $e$.  We will therefore refer to such engines as examples of ``chemical engines'', even though the transfer of charge associated with their operation is not a chemical reaction in the usual sense.\footnote{Chemical reactions depend on the transfer of electrons.  The widespread use of ``electrochemical potentials'', which combine a potential associated with a chemical reaction and an ``electrostatic'' contribution $eV$, underlines the fact that there is no sharp thermodynamic distinction between electron transfer proccesses associated with chemical changes and those that are not.} Although macroscopic electrostatic engines are much less powerful than electromagnetic engines, we will focus on the former here both because their feedback dynamics is simpler to model mathematically and also because they are much easier to miniaturize, which suggests that they may be of considerable practical importance in biophysics and other mesoscopic applications \cite{LEC}.


\subsubsection{Franklin bells}
\la{sec:bells}

\begin{figure} [t]
\begin{center}
	\includegraphics[width=0.35 \textwidth]{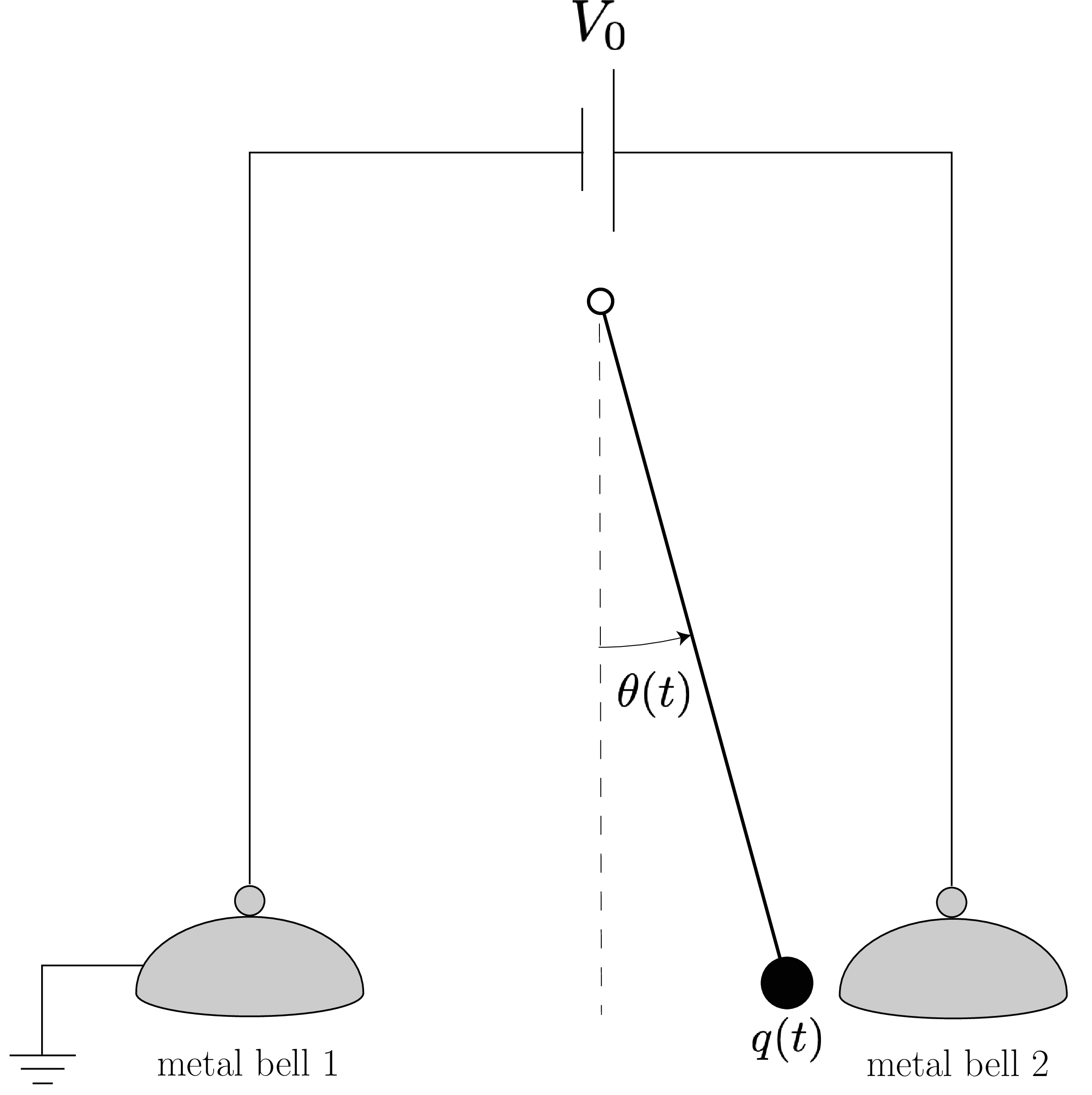}
\end{center}
\caption{\small Franklin bells: the self-oscillation of the pendulum is powered by the electrostatic potential difference $V_0$ between the two metal bells.  The active force the drives pendulum results from a positive feedback between the angular position of the pendulum, $\theta(t)$, and the rate of change of its charge $q(t)$.\la{fig:bells}}
\end{figure}

The scheme of the ``Franklin bells'' is shown in \Fig{fig:bells}.  This admits a very similar dynamical description to the putt-putt heat engine of \Sec{sec:active}.  For a pendulum of length $\ell$ and mass $m$, with vertical gravitational acceleration $g$, and with the angle $\theta$ restricted to small values ($ | \theta | \ll 1$), we have
\be
\ddot \theta  + \gamma \dot \theta + \frac g \ell \theta = \frac{q}{m \ell^2} E_{\rm h}(\theta)
\la{eq:bell-theta}
\ee
where $E_{\rm h} (\theta)$ is the value of the horizontal component of the electric field between the bells, at the position corresponding to the pendulum with angular cooordinate $\theta$.   For simplicity, we take this as constant:
\be
E_{\rm h} = \frac{V_0}{2 \ell \theta_0}
\ee
where $\pm \theta_0$ are the positions at which the pendulum strikes the bells.  By Ohm's law, the charge on the pendulum varies as
\be
\dot q = -\frac{V(q, \theta)}{R_1 (\theta)} + \frac{V_0 - V(q, \theta)}{R_2 (\theta)}
\la{eq:bell-q}
\ee
where $V(q, \theta)$ is the voltage of the pendulum with charge $q$ at position $\theta$, while $R_{1,2} (\theta)$ are the resistances for current flowing from the pendulum to the respective bells, with $R_1(\theta)$ an increasing function and $R_2(\theta)$ a decreasing one.  (In practice, we expect each of these resistances to fall off abruptly when the pendulum approaches the corresponding bell.)  If $C$ is the capacitance of the charged pendulum bob, we have that
\be
V(q, \theta) = \frac{q}{C} + \frac{V_0 (1 - \theta/\theta_0)}{2} .
\ee
Equation \eqref{eq:bell-q} can therefore be expressed in same form as \Eq{eq:N}, as
\be
\dot q = - D(\theta) q + C(\theta) 
\ee
and the conditions for self-oscillation can be found by the same sort of analysis discussed in \Sec{sec:active}.

We will not work out here further details of the dynamics of the Franklin bells, except to note that the active force results, as in the putt-putt engine, from the positive feedback between the mechanical degree of freedom of the tool (in this case, the pendulum's $\theta$) and the rate of change of another variable the controls the force acting on the tool (in this case, the electric charge $q$).  In terms of the Rayleigh-Eddington criterion, the Franklin bells work as an engine because the pendulum absorbs matter (charge) at higher chemical potential (corresponding to the higher voltage of bell 2 in \Fig{fig:bells}) and rejects it at a lower chemical potential (corresponding to the lower voltage of bell 1).  Its operation therefore agrees entirely with the general thermodynamic principles discussed in \Eq{eq:mu-modulation}.

\begin{figure} [t]
\begin{center}
	\includegraphics[width=0.4 \textwidth]{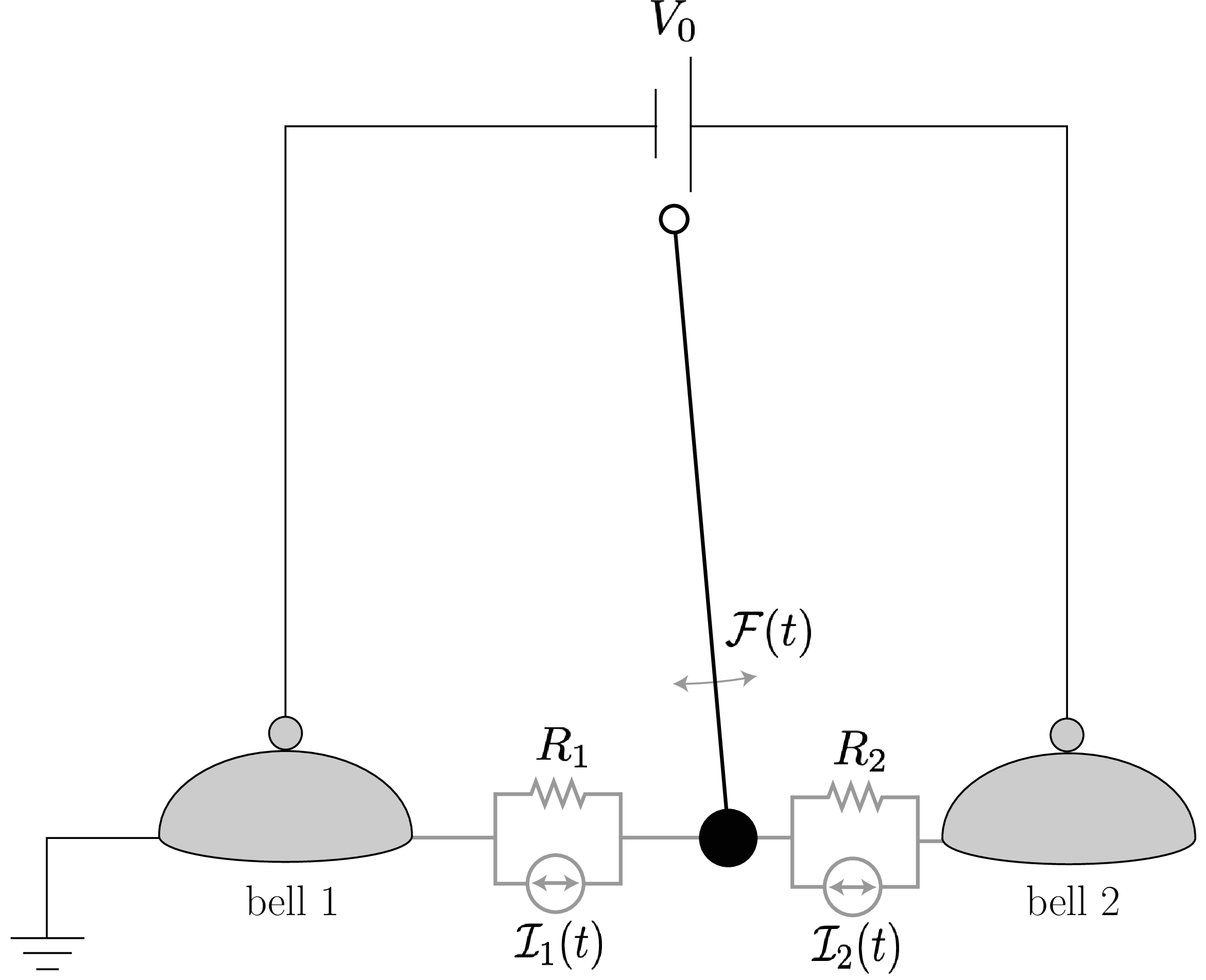}
\end{center}
\caption{\small Scheme for incorporating thermal noise into the equations of motion for the Franklin bells.  The stochastic force ${\cal F}(t)$ is associated, by a fluctuation-dissipation relation, to the mechanical damping exerted on the pendulum's motion by its medium, as represented by the $\gamma \dot \theta$ term in \Eq{eq:bell-theta}.  The stochastic currents ${\cal I}_{1,2}(t)$ correspond to the Johnson-Nyquist noise associated with the corresponding resistances $R_{1,2}$ in \Eq{eq:bell-q}.\la{fig:bells-noise}}
\end{figure}

So far, our treatment of the Franklin bells has been deterministic, which is appropriate for a macroscopic system.  The effects of thermal noise can be introduced into the equations of motion by adding to \Eq{eq:bell-theta} the stochastic term associated to the damping $\gamma \dot \theta$ by the fluctuation-dissipation relation, and to \Eq{eq:bell-q} the Johnson-Nyquist noise associated with the variable resistances $R_{1,2} (\theta)$ (see, e.g., \cite{Pathria}).  This is illustrated in \Fig{fig:bells-noise}.  In \Sec{sec:stochastic-osc} we will see how this translates to the description of the nanoscopic ``electron shuttle'' proposed in \cite{shuttle}.


\subsubsection{Quincke rotor}
\la{sec:Quincke}

In 1893, Weiler reported that a glass cylinder immersed in a conducting fluid rotated spontaneously if a large enough electrostatic field was applied to the fluid \cite{Weiler}.  A similar phenomenon was discovered independently by Quincke in 1896, using small solid spherical particles \cite{Quincke}.  This phenomenon is therefore commonly known as ``Quincke rotation''.  For a modern review of the experimental and theoretical literature on this subject, see \cite{Jones}.

Here we formulate our own theoretical treatment of the dynamics of a simple Quincke rotor, which will serve as a paradigm for other engines with rotating tools.  As we shall see, since the sign of the angular velocity $\omega$ has a physical meaning for a rotor that it lacks for a one-dimensional oscillator (see \Sec{sec:RE}), the action of the external load on the tool's mechanical degree of freedom is particularly simple to represent for a rotating engine.  Our model consists of a dielectric cylinder of the radius $R$ immersed in a conducting fluid with a uniform electric field $\vv E$ through it, as shown in \Fig{fig:quincke}. The cylinder has length $L$ and can spin about a fixed axis.  The current density $\vv j$ through the fluid is given by Ohm's law:
\be
\vv j = \sigma \vv E ,
\ee
where $\sigma$ is the fluid's conductivity.  We assume that the system is uniform in the axial direction, leaving the angle $\phi$ as the only relevant mechanical degree of freedom.  We treat $\phi$ as an unbounded real number, from which we can deduce the number of net turnings of the cylinder, relative to its initial state.  In terms of the angular velocity $\omega \equiv \dot \phi$, the rotor's equation of motion takes the form
\be
\dot{\omega} + \gamma \omega = \tau - \tau_{\rm load}
\la{eq:eq_motion}
\ee
where $\tau$ is a torque (divided by the rotor's moment of inertia) produced by the electrostatic force and $\tau_{\rm load}$ accounts for the attached load. The accumulated surface charge $\rho(t, \phi)$ depends only on the angle (modulo $2\pi$) and time, such that $\rho(t, \phi) \, d\phi$ is equal to the charge on the strip $L R\, d\phi$ along the cylinder's surface. Charge conservation implies that
\be
\frac{\partial}{\partial t} \rho(t, \phi) = -\omega(t) \frac{\partial}{\partial \phi} \rho(t, \phi) + L R \left[ \sigma E \cos \phi - \frac{\sigma}{4\pi \epsilon} \rho(t, \phi) \right] ,
\la{eq:charge_dyneq}
\ee
where $\epsilon$ is the fluid's dielectric constant.  The first term in the right-hand side of \Eq{eq:charge_dyneq} accounts for the cylinder's rotation.  The second term describes the deposition of charge upon the cylinder's surface by the current $\vv j$ through the fluid.  The third term describes the radial outflow of charge due to electrostatic repulsion, with
\be
j_{\rm out} = \frac{L R \sigma}{4 \pi \epsilon} \rho (t, \phi)
\ee
(see \Fig{fig:quincke}).  We have assumed that the conductivity of the cylinder is much smaller than $\sigma$, so that we may neglect any currents within the cylinder or along its surface.

\begin{figure} [t]
\begin{center}
	\includegraphics[width=0.38 \textwidth]{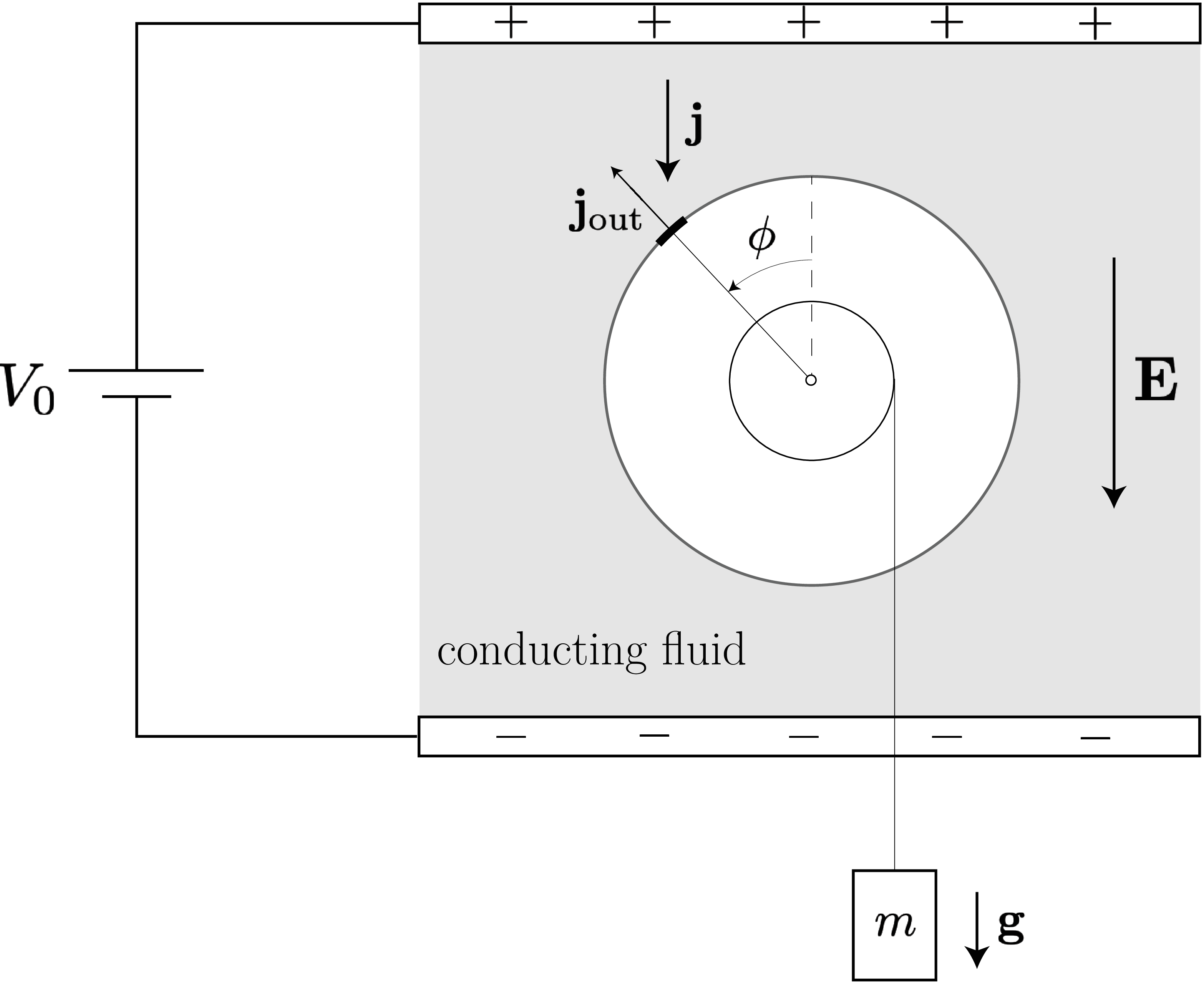}
\end{center}
\caption{\small Quincke rotor: the self-rotation of the dielectric cylinder is powered by the electrostatic potential difference $V_0$ between the parallel plates.  The torque results from a positive feedback between the turning of the cylinder, described by the angle $\phi$, and the change in the charge attached to the cylinder's surface.  This torque can lift a weight $m$, which acts as the load for the engine.  The weight and the string attached to it are assumed to be uncharged.\la{fig:quincke}}
\end{figure}

The normalized torque produced by the charge distribution $ \rho(t, \phi)$ is
\be
\tau(t) = \frac{L R E}{I}\int_0^{2\pi} d\phi\, \sin\phi \rho(\phi;t)
\la{eq:torque}
\ee
where $I$ is the momentum of inertia of the cylinder.  Since $\rho(t, \phi)$ is $2\pi$-periodic function of $\phi$, it can be decomposed into Fourier components 
\be
\rho(t, \phi) = a_0 (t) + a_1 (t) \cos \phi + b_1(t) \sin \phi + a_2(t) \cos 2\phi + b_2(t) \sin 2\phi + \ldots
\la{eq:fourier}
\ee
Combining \Eq{eq:fourier} with Eqs.\ \eqref{eq:eq_motion}, \eqref{eq:charge_dyneq}, and \eqref{eq:torque}, we obtain three coupled equations of motion for $\omega(t)$, $a_1(t)$, and $b_1(t)$:
\bea
&& \dot \omega + \gamma \omega =\frac{L R E}{2I} b_1 - \tau_{\rm load} , \la{eq:dyn_omega} \\
&& \dot a_1 + \frac{L R \sigma}{4\pi\epsilon} a_1 + \omega b_1 =  L R \sigma E , \la{eq:dyn_a} \\
&& \dot b_1 + \frac{L R \sigma}{4\pi\epsilon} b_1 - \omega a_1 =  0 . \la{eq:dyn_b}
\eea
The other Fourier components are damped and undriven, and therefore relax to zero.  Introducing the short-hand notation  
\be
\Gamma \equiv \frac{L R \sigma}{4\pi\epsilon} , \quad \mu \equiv \frac{L R E}{2I} , \quad J \equiv L R \sigma E ,
\la{eq:notation}
\ee
and the new complex dynamical variable
\be
z(t) \equiv a_1(t) + i b_1(t) ,
\la{eq:zet}
\ee
we can recast Eqs.\ \eqref{eq:dyn_omega}, \eqref{eq:dyn_a}, and \eqref{eq:dyn_b} into the form 
\be
\left\{ \begin{array}{l}
\displaystyle \dot{\omega} = - \gamma\omega +\frac{\mu}{2i} (z - \bar z) - \tau_{\rm load} \\
\displaystyle \dot{z} = (i\omega -\Gamma )z + J \phantom{+ \frac 0 1} \end{array} \right. ,
\la{eq:dynsystem1}
\ee
where $\bar z \equiv a_1 - i b_1$ is the complex conjugate of $z$.  This describes a rotor coupled to a damped oscillator fed by the external energy source and loaded by a constant external torque.  The coupling between $\omega$ and $z$ corresponds to the feedback that produces the active torque that drives the engine.  Note that this feedback involves a nonlinearity (the $i \omega z$ term in \Eq{eq:dynsystem1}) and is therefore qualitatively different from the linear feedback that we considered before in \Eq{eq:matrix}.

It is easy to show that, in the absence of a load ($\tau_{\rm load} = 0$), \Eq{eq:dynsystem1} has steady self-rotating solutions ($\dot \omega = \dot z = 0$ with $\omega \neq 0$) for $\mu J > \gamma \Gamma^2$, which can be translated into the minimum electrostatic field needed for a cylinder of fixed dimensions and moment of inertia to start spinning.  The corresponding steady rotation has angular velocity
\be
\omega_{\rm st} = \pm \sqrt{ \frac{\mu J}{\gamma}  - \Gamma^2} ~.
\la{eq:wst}
\ee
In the mathematical theory of dynamical systems, this is classified as a pitchfork bifurcation \cite{Strogatz}.

We will not attempt a full characterization of the phases of \Eq{eq:dynsystem1} as a dynamical system, which is an interesting problem in its own right.  Numerical simulations show that it has a rich behavior, which includes stable, self-rotating, and chaotic regimes.  Thermal noise can be incorporated into the dynamics of this Quincke rotor model by including stochastic terms for the mechanical degree of freedom (\Eq{eq:zet}).  We will return to this in \Sec{sec:stochastic-rot}, after we consider how to extend the Fokker-Planck and Langevin equations to simple engines.


\section{Fluctuations}
\la{sec:fluctuations}

The deterministic equations of motion that we considered in \Sec{sec:feedback} are valid for macroscopic engines with heavy tools, but as the engine is miniaturized the thermal noise becomes more important.  This noise is unavoidable, since the active force that drives the engine's tool is always associated with dissipation.  The thermal fluctuations therefore impose limitations on how far an engine can be miniaturized before its operation becomes too erratic or fails altogether.  To understand these fluctuations we must describe the engine's active dynamics in terms of statistical physics.

In this section we will present a dynamical treatment of an engine as an open system, based on the master-equation formalism.  Such a treatment for a heat engine was first proposed in \cite{Alicki1979, Kosloff1984}, and it has more recently been refined and extended in the context of ``quantum  thermodynamics''; see \cite{QT} and references therein.  This section is intended as a review of the subject from an up-to-date perspective, applicable to both heat and chemical engines.  Our goal is to introduce the key concepts and formalisms to help the non-specialist understand the original conceptualization and results that will then be developed in \Sec{sec:dynamics}.

As we mentioned in \Sec{sec:intro}, within the specific context of adiabatic quantum motors driven by current-induced forces (which, in our nomenclature, are rotating chemical engines) there is an alternative dynamical treatment in the literature, based on the scattering matrix for the conductor and the Keldysh formalism for the corresponding non-equilibrium Green's functions; see especially \cite{wheel, Oppen}, in addition to the references \cite{current, AQM1, AQM2} already mentioned in the Introduction.  Although we expect that our principal results may be translated into this scattering formalism, we prefer the MME approach because it seems to us more flexible and general, as well as more transparently connected with statistical mechanics.

\subsection{Quantum thermodynamics}
\la{sec:QT}

The development of a mathematically consistent theory of open quantum systems offers new tools for the derivation, from first principles, of dynamical models for heat and chemical engines \cite{QT}.  Paradoxically, quantum models with discrete energy microstates (but admitting continuous reversible dynamics) are often more tractable than classical models based on a continuous phase-space.  In particular, the quantum working substance can be modeled either by a system with a finite-dimensional Hilbert space (even a two-state system will do!) or by a few harmonic oscillators.  For such finite quantum systems interacting with several heat and/or chemical baths in the weak-coupling (or, alternatively, low-density) regime, one can derive the  equation of motion for its reduced density matrix $\rho(t)$ \cite{AL}.  This takes the form of a Markovian Master Equation (MME), which in the Schr\"odinger picture can be expressed as:
\be
\dot{\rho}(t) = - i [ H (t) , \rho(t) ] + \sum_k \La_k (t) \rho(t) ,
\la{eq:MME}
\ee
where $H(t)$ is the system Hamiltonian.  The $\La_k$'s, known as {\it dissipators}, are mathematical superoperators that describe the non-unitary evolution of the state $\rho(t)$ of the system due to the loss of information about its entanglement with environment.  We work throughout in units such that $\hbar = 1$.  The Markovian character of \Eq{eq:MME} (meaning that $\dot \rho (t)$ depends only on $\rho (t)$ and not on its entire history) is an approximation valid when the correlations in the environment decay sufficiently quickly, allowing us to treat the environment as a collection of simple thermodynamic baths \cite{AL}.  We also require that the external modulation of $H(t)$ be slow compared to the time scales that control the system's internal dynamics.

Equation \eqref{eq:MME} can be re-expressed in the Heisenberg picture (with the time dependence in the operator $A$ rather than the state $\rho$) as:
\be
\dot{A}(t) = i [H (t) , A(t)] +  \sum_{k} \La^\ast_k(t) A(t) ,
\la{eq:MMEh}
\ee
with the defining relation  
\be
\Tr[\rho(t) A] = \Tr[\rho A(t)] .
\la{eq:Heisenberg}
\ee
The time-dependence of the system Hamiltonian $H(t)$ accounts for the cyclic changes of external conditions caused by the motion of the macroscopic tool. This induces a time-dependence in each of the dissipators $\La_k (t)$ that describe the influence of the $k$-th bath on the working substance. Each bath is assumed to be a large quantum system characterized by inverse temperature $\beta_k$ and chemical potential $\mu_k$.  If the working substance can exchange with the bath (quasi)particles that satisfy a conservation law, the particle number operator $N$ (with $[H(t), N] = 0$) becomes relevant. In such situations the dissipator $\La_k(t)$ drives the system into the equilibrium state
\be
\rho^{\rm eq}_k(t)  = Z(t)^{-1} e^{- \beta_k [ H(t) - \mu_k N ] } \quad \hbox{with} \quad  \La_k(t) \rho^{\rm eq}_k(t) = 0 
\la{eq:eqstate}
\ee
(i.e., the Gibbs state corresponding to that bath).

To show that this model is consistent with the laws of thermodynamics, we first propose the natural microscopic definitions of internal energy $\Ua(t)$ and entropy $\Sa(t)$.  In units with $k_B \equiv 1$,
\be
\Ua(t) \equiv \langle H(t)\rangle_{\rho} \quad \hbox{and} \quad \Sa(t) \equiv \langle \ln\rho(t)\rangle_{\rho}
\la{eq:US}
\ee
expressed in a shorthand notation for the average with respect to the temporal state of the system:
\be
\langle A \rangle_{\rho}\equiv \Tr [\rho(t) A]  .
\la{eq:average}
\ee
We will use the following \emph{Spohn inequality} \cite{Austin, Spohn}, valid for any dissipator $\La$ possessing a stationary state $\rho_0$ (i.e. $\La \rho_0 = 0$):
\be
\langle \La^\ast (\ln \rho - \ln \rho_0)\rangle_\rho \leq 0  .
\la{eq:Spohn}
\ee
The entropy balance for working substance is then
\bea
\dot \Sa (t) &=& - \Tr[\dot \rho (t) \ln \rho(t)] = -\sum_k \left\langle \La_k^\ast (t) \ln \rho(t) \right\rangle_\rho \nl 
 &=& -\sum_k \left\langle \La_k^\ast (t) \left[ \ln \rho(t) - \ln \rho^{\rm eq}_k(t) \right] \right\rangle_\rho + \sum_k \beta_k \left\langle \La_k^\ast(t)  \left[ H(t) - \mu_k N \right] \right\rangle_\rho .
\la{eq:entropybalance}
\eea 
Defining the heat current flowing from the $k$-th baths as
\be
\Ja_k (t) = \left\langle \La_k^\ast (t)  [ H(t) - \mu_k N ] \right\rangle_\rho
\la{eq:heatcurrent-q}
\ee
and using \Eq{eq:Spohn}, one can rewrite the entropy balance of \Eq{eq:entropybalance} as the second law of thermodynamics:
\be
\dot \Sa (t) - \sum_k \beta_k \Ja_k (t) \geq 0 .
\la{eq:2law}
\ee
Similarly, the energy balance expressed in the form
\be
\dot \Ua (t) =  \left\langle \dot H(t) \right\rangle_\rho + \sum_k \left\langle \La_k^\ast (t) H(t) \right\rangle_\rho
\la{eq:energybalance}
\ee 
can be rewritten as the first law of thermodynamics,
\be
\dot \Ua (t) = \Ja (t)  + \sum_k \mu_k \dot \Na_k (t) - \Pa(t) , 
\la{eq:1law}
\ee 
where 
\be
\Ja (t)  = \sum_k \Ja_k (t)
\la{eq:totalheat}
\ee
is the total heat current flowing from the baths, while
\be
\dot \Na_k (t) = \left\langle \La_k^\ast(t) N \right\rangle_\rho
\la{eq:particle_current}
\ee 
is a particle current flowing from the $k$-th bath, and
\be
\Pa(t) = - \left\langle \dot H (t) \right\rangle_\rho
\la{eq:P}
\ee 
is the power supplied by the working substance to the tool.

The classical thermodynamic analysis of \Sec{sec:RE} had already taught us that the macroscopic state of the working substance must vary with time in order for an engine to produce net positive work.  The lesson of the quantum thermodynamic analysis in this section is complimentary to that: in order  to define the engine's power output $\cal P$ in \Eq{eq:1law}, we had to give the working substance a time-dependent Hamiltonian $H(t)$.  This time dependence results from the motion of a macroscopic or mesoscopic degree of freedom that interacts with the working substance.  In an autonomous engine, that macrosopic or mesoscopic degree of freedom is the tool, and work results from the positive feedback between the state of the working substance and the mechanical state of the tool.  This feedback produces the active force that drives the tool's cyclical motion.   Although the feedback needed to generate that active force is not manifest in the approach that we have sketched in this section, the presence of an active force can be deduced from the positive power output when \Eq{eq:1law} is averaged over a complete cycle.

\subsection{Stochastic thermodynamics}
\la{sec:ST}

A full dynamical description of an autonomous engine that includes the feedback mechanism and a physically acceptable modeling of the load is difficult to formulate as a quantum model because of the well known problems with the validity of the Markovian approximation for open quantum systems composed of several interacting subsystems whose evolution is governed by different time scales.  Note that our conceptualization of an autonomous engine includes three components: the working substance, the tool, and the load, all of which interact with the non-equilibrium environment composed of baths with different temperatures and/or chemical potentials.  Previous theoretical work on this subject in the context of quantum thermodynamics has not distinguished between the tool and the load, instead modeling the tool as a flywheel that accumulates energy during some brief time of initial operation.  Numerical results show an approximately linear initial energy increase, from which the engine's power output has been computed by subtracting the ``passive energy'' that heats the flywheel \cite{q-amps, flywheel, autorotor}.

Our aim here is to describe autonomous engines permanently attached to the load and finally running in the steady-state regime.  For this we will use a classical stochastic model, which is easier to construct but is still applicable to mesoscopic systems in which thermal fluctuations are relevant.   We can do this in a formalism that is similar to the derivation of the quantum MME sketched above.  The density matrix $\rho(t)$ is replaced by the probability distribution $\vv p (t) =  p(t,x)$ where $x$ is a point in phase-space that describes the relevant system and which can combine discrete and continuous variables. The average value of the observable $A(x)$ is denoted by
\be
\langle A \rangle_{\vv p} \equiv \int dx\, A(x) p(t, x) ,
\la{eq:A-ex}
\ee
where $\int dx$ means integration over continuous variables and summation over discrete ones. The Markovian evolution equation can be expressed as
\be
\dot{\vv p}(t) =  \sum_k \La_k \mathbf{p}(t) +  \La_{\rm work} \vv p (t)  ,  
\la{eq:MME_cl}
\ee
where all dissipators are time-independent (autonomous model). The dissipator $ \La_k $ describes the influence of the $k$-th equilibrium bath and possesses the form of a Pauli master equation for discrete variables, and of a linear Boltzmann equation or a Fokker-Planck equation for continuous variables. The time-independent version of the condition of \Eq{eq:eqstate} is satisfied by stationary states of the form
\be
 \La_k \vv p^{\rm eq}_k = 0, \quad \vv p^{\rm eq}_k  = Z^{-1} e^{- \beta_k [ H - \mu_k N ]} ,
\la{eq:eqstate_cl}
\ee
where now $H \equiv H(x)$ is the energy and $N \equiv N(x)$ is the particle number in the microstate $x$.  We can define ``Heisenberg-picture'' dissipators $\La^\ast_k$, by analogy to the quantum analysis of \Sec{sec:QT}, by having the dissipator act on $A$ rather than $p$ in \Eq{eq:A-ex}.  For example, for a differential operator $\La_k$ we obtain $\La^\ast_k$ by integrating by parts.

The dissipator $\La_{\rm work}$ takes into account the energy transfer to the load, which should correspond to the engine's output work. Therefore, this energy transfer cannot be accompanied by any entropy change, implying that the equality
\be
\langle  \La^\ast_{\rm work} \ln \vv p \rangle_{\vv p} = 0   
\la{eq:workdiss}
\ee
is satisfied for arbitrary $\vv p$.

The classical analog of the Spohn inequality (\Eq{eq:Spohn}) allows us to express the laws of thermodynamics as: 
\bea
\hbox{First law:} &\quad& \dot{\Ua} = \Ja + \sum_k \mu_k \dot \Na_k - \Pa , \la{eq:1law_cl} \\
\hbox{Second law:} &\quad& \dot{\Sa}(t) - \sum_{k} \beta_k \Ja_k (t) \geq 0 . \la{eq:2law_cl}
\eea
The internal energy is given by
\be
\Ua = \langle H \rangle_{\vv p}
\la{eq:defU_cl}
\ee 
and the output power is defined as
\be
\Pa = -\langle \La^\ast_{\rm work} H \rangle_{\vv p}.
\la{eq:defP_cl}
\ee 
The heat currents are
\be
\Ja_k (t) = \langle \La_k^\ast  [H - \mu_k N]\rangle_{\vv p}.
\la{eq:heatcurrent-cl}
\ee

Here we have translated the quantum-thermodynamic MME of \Sec{sec:QT} to a classical statistical description.  The resulting formalism is similar to the ``stochastic thermodynamics'' approach that has been developed independently by others (see \cite{Seifert} and references therein).  Unlike in the usual stochastic thermodynamics treatment, however, we will not be concerned here with calculating entropy production over specific trajectories.  Moreover, we seek to describe the active force that drives the engine in terms of a realistic and autonomous feedback dynamics, rather than as an external driving or an anti-damping.
 

\section{Deterministic and stochastic engine dynamics}
\la{sec:dynamics}

Let us now consider some examples of autonomous engines, both in the deterministic macroscopic regime and in their corresponding mesoscopic versions that incorporate thermal noise.  We will treat both self-oscillating and self-rotating engines.


\subsection{Deterministic model of oscillating engine}
\la{sec:deterministic}

We may generalize the putt-putt model of \Sec{sec:active} and the Franklin bells model of \Sec{sec:bells} into a generic dynamical system described by coupled homogenous equations of motion of the form
\bea
\dot x &=& v , \la{eq:dyneq-x} \\
\dot v &=& - \Omega_0^2 x - \gamma v - a n , \la{eq:dyneq-v} \\
\dot n &=& \gamma_+ (x)  - \gamma_- (x) n \la{eq:dyneq-n} .
\eea
Equations \eqref{eq:dyneq-x} and \eqref{eq:dyneq-v} describe a harmonic oscillator (the tool) with damping rate $\gamma$, subject also to a force proportional to $n$, which is exerted by the working substance.  Equation \eqref{eq:dyneq-n} is the kinetic equation for the quantity of working substance, where the production rate $\gamma_+ (x)$ and the decay rate $\gamma_-(x)$ depend on the position $x$ of the tool.  The quantity $n$ of this substance is computed with respect to a certain reference value, such that the fixed point of Eqs.\ \eqref{eq:dyneq-x}, \eqref{eq:dyneq-v}, and \eqref{eq:dyneq-n} is given by $x = n = 0$ and $\gamma_+(0)=0$.  Note that \Eq{eq:dyneq-n}, which is necessary for the feedback that can give rise to an active force, describes an {\it irreversible} dynamics.  This can be seen mathematically in that \Eq{eq:dyneq-n} is not invariant under the $t \to -t$ transformation, and also in that the physical processes that it may represent (such as the condensation and evaporation of water in the putt-putt engine or the Ohmic conduction in the Franklin bells) are thermodynamically irreversible processes that break detailed balance.

The mechanical damping constant $\gamma$ in \Eq{eq:dyneq-v} can be decomposed into two thermodynamically distinct contributions,
\be
\gamma = \gamma_{\rm diss} + \gamma_{\rm load}
\la{eq:gamma}
\ee
where $\gamma_{\rm diss}$ describes the ambient friction acting on the oscillator, while $\gamma_{\rm load}$ accounts consumption of work by the load attached to it.  This $\gamma_{\rm load}$ will be needed to the unambiguous identification of the work output from the engine, as explained  in \Sec{sec:ST}.

Rather than repeat the same stability analysis used in \Sec{sec:active} (or to extend it to the nonlinear regime in order to find limit cycles or strange attractors as in \cite{LEC}), let us here use a different approach that may help to illuminate the nature of the non-conservative force that arises from the feedback between $x$ and $n$.  For simplicity, we begin by linearizing Eqs.\ \eqref{eq:dyneq-x}, \eqref{eq:dyneq-v}, and \eqref{eq:dyneq-n} to obtain
\be
\left\{ \begin{array}{l}
\displaystyle \dot x = v , \\ 
\displaystyle \dot v =  - \Omega_0^2 x - \gamma v - a n , \\
\displaystyle \dot n = b x - \Gamma n \end{array} \right.
\la{eq:dyneq_lin}
\ee 
with 
\be
\Gamma \equiv \gamma_- (0), \quad b = \left. \frac{d \gamma_+}{dx} \right|_{x=0}.
\la{eq:dyneq_lin1}
\ee
Integrating the third relation in \Eq{eq:dyneq_lin} for initial value $n(0) = 0$, we obtain the equation of motion for the tool degree of freedom in the Newtonian form
\be
\ddot x + \gamma \dot x + \Omega_0^2 x = - ab \int_0^t ds \, e^{-\Gamma s} x(t - s) .
\la{eq:Newton}
\ee
The right-hand side of \Eq{eq:Newton} corresponds to the active force (per unit mass).  The product $ab$ is evidently a measure of the strength of the feedback between $x$ and $n$ in \Eq{eq:dyneq_lin}.  Finnie and Curl wrote such an integro-differential equation of motion for their model of the putt-putt boat in \cite{pp2}, but this approach has not, as far as we know, been used elsewhere in either the mathematical or the physical literature.  Equation \eqref{eq:Newton} clearly shows that the active force is not given by the gradient of any potential function, not even a time-dependent one.

Note that, for large $\Gamma$, the Taylor expansion of $x(t-s)$ to first order in $s$ in the right-hand side of \Eq{eq:Newton} gives an effective negative damping, as in the Rayleigh, van der Pol, and Stuart-Landau models for one-dimensional self-oscillators.  On the other hand, if we assume that $x(t)$ varies with a period that is very long compared to $1/\Gamma$, the integral on the right-hand side of \Eq{eq:Newton} can be replaced by a retarded force proportional to $x(t-c)$, for the appropriate choice of a fixed $c$.  This helps explain the successful application of phenomenological models of one-dimensional self-oscillators that use negative damping or finite delays, rather than a physically realistic description of the positive feedback.  The Laplace transform method can be applied to \Eq{eq:Newton} to reduce it to an algebraic equation, facilitating the calculation of the conditions for self-oscillation.


\subsection{Stochastic model of oscillating engine}
\la{sec:stochastic-osc}

Let now now consider the ``electron shuttle'' proposed in \cite{shuttle}.  This is essentially a nanoscopic version of the Franklin bells that we considered in \Sec{sec:bells}.  In our analysis of the electron shuttle we will use similar stochastic evolution equations to those formulated in \cite{stochastic-shuttle}, but our discussion of the energy and entropy balances is different and consistent with the approach that we have outlined in \Sec{sec:QT}.

\begin{figure*} [t]
\begin{center}
	\subfigure[]{\includegraphics[height=0.28 \textwidth]{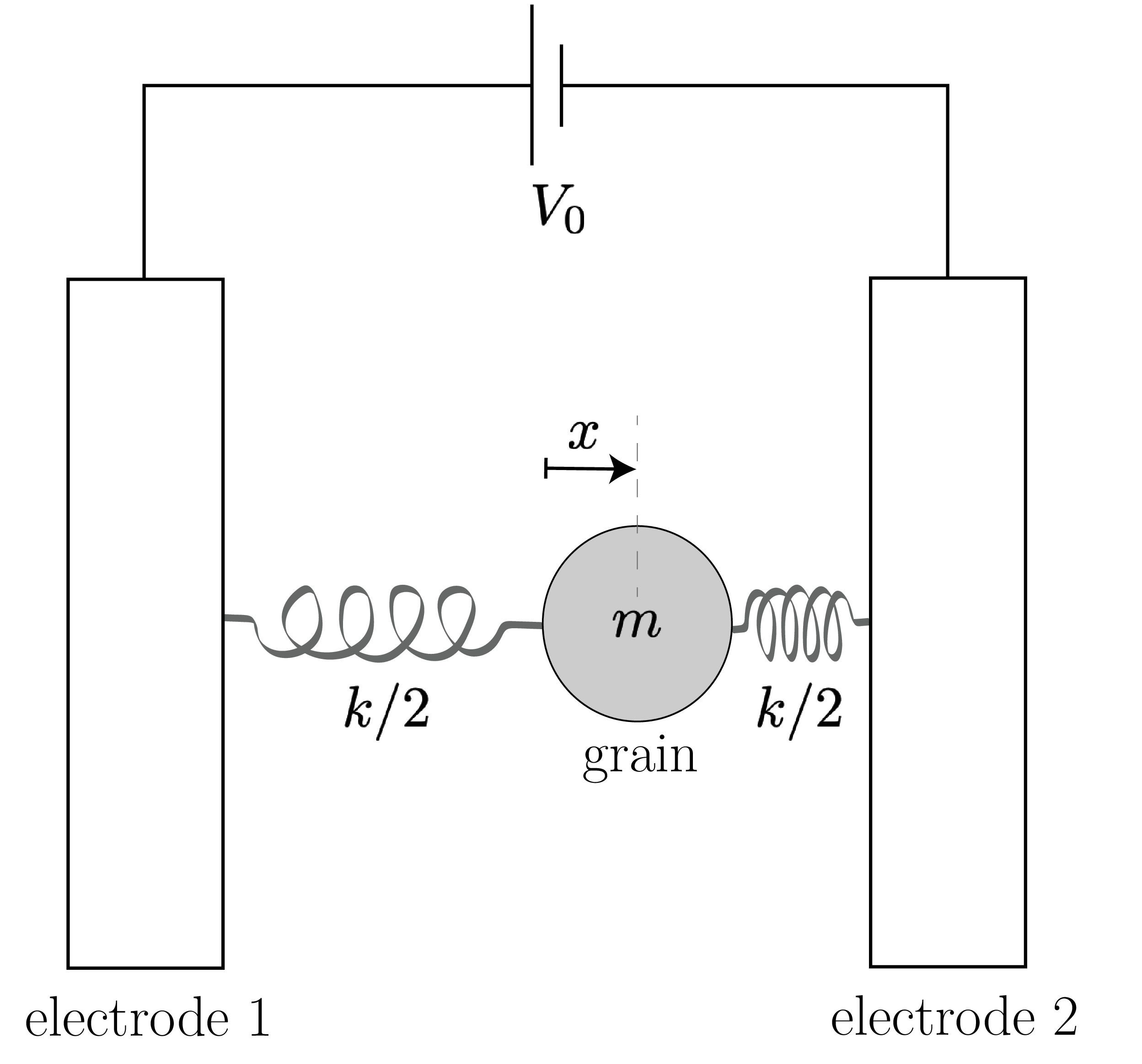}} \hskip 60 pt
	\subfigure[]{\includegraphics[height=0.28 \textwidth]{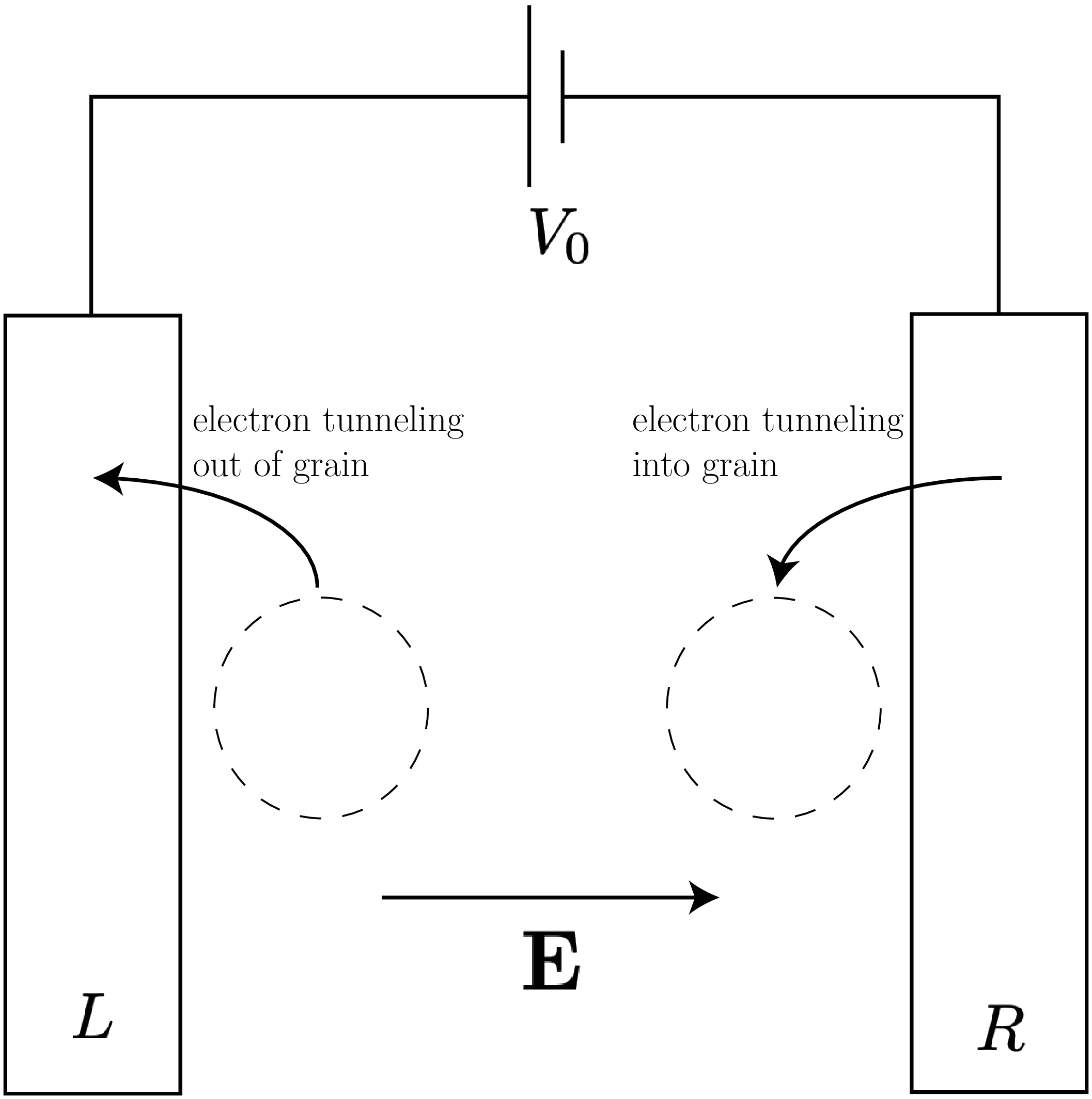}}
\end{center}
\caption{\small Electron shuttle: (a) A metallic grain of mass $m$ is mounted on a harmonic oscillator with elastic constant $k$.  The position of the grain, measured with respect to its equilibrium, is denoted by $x$. The oscillator is placed in a constant electric field $\vv E$ maintained by a voltage source $V_0$ connected to two electrodes.  (b) Electrons can tunnel between the grain and the electrodes, changing the net charge upon which the field $\vv E$ acts.  The electrodes may be considered as thermodynamic reservoirs $L$ and $R$.  Images adapted from \cite{shuttle}.\la{fig:shuttle}}
\end{figure*}

As shown in \Fig{fig:shuttle}, the electron shuttle consists of a metallic grain of mass $m$ mounted on a mechanical oscillator with elastic constant $k$.  This oscillating grain acts the engine's tool (piston).  The energy of its self-oscillation can be fully dissipated (as would be the case if the shuttle acted as a clock) or used in part to drive an external load.  In our analysis, we will consider the grain as a quantum dot with a single state of the energy $\epsilon$ that can be occupied by a single electron. This electron in the dot constitutes the engine's working substance, fully characterized by the occupation number  $N = 0, 1$. The total energy for the dot-oscillator system reads
\be
U (x,v,N) =  \frac 1 2 \left( v^2 + \Omega_0^2 x^2 \right) + \epsilon N +  a N x .
\la{eq:Ushuttle}
\ee
where $\Omega_0^2 = k / m$ and $a = e E$, where $e$ is the elementary charge and $E$ the magnitude of the constant electric field between the two electrodes.

The electron in the dot is an open quantum system, weakly coupled to two baths corresponding to the bulk materials $L$ and $R$ at equilibrium states characterized, in general, by different inverse temperatures $\beta_X$ and chemical potentials $\mu_X$,  ($X= L,R$). The standard derivation gives us the stochastic equation for the probabilities $p(N)$ (with $p(0) = 1 - p(1)$):
\be
\dot p(1) = \gu(x)p(0) - \gd(x) p(1) = \gu(x) - \left[  \gd(x)  +  \gu(x) \right] p(1) ,
\la{eq:dotstoch}
\ee
with the decay rate $\gd(x)$  and the pumping rate $\gu(x)$ depending on the position of the piston $x$.  Since electron transport involves tunneling between the dot and a reservoirs, we can decompose the rates as 
\be
\gu(x) = \gu^{L}(x) + \gu^{R}(x), \quad \gd(x) =  \gd^{L}(x) + \gd^{R}(x) ,
\la{eq:guX}
\ee
satisfying the Kubo-Martin-Schwinger (KMS) relation 
\be
\gu^{X}(x) =  e^{-\beta_X(\epsilon + a x - \mu_X)} \gd^{X}(x), \quad X= L, R .
\la{eq:KMS}
\ee
Note that for our single-electron dot the probability $p(1)$ is equal to the average number of electrons in the dot, say $n$.  The stochastic \Eq{eq:dotstoch} is then mathematically equivalent to the deterministic kinetic equation for the amount of working substance in \Eq{eq:dyneq-n}.   Similarly, the deterministic \Eq{eq:dyneq-n} can be transformed first into a Langevin equation by adding the thermal noise, and finally into a Fokker-Planck equation for the oscillator. Therefore, the time evolution of the joint probability distribution $p(x, v, N; t)$ is governed by the Fokker-Planck equation for continuous variables $(x,v)$, combined with the MME for the discrete space $N = \{ 0,1\}$:
\bea
\frac{\partial}{\partial t} p(x, v, N; t) &=& - v\frac{\partial}{\partial x} p(x, v, N; t) + \left[ \frac{\partial}{\partial v} \left(\Omega_0^2 x + a N +\gamma v\right) + D\frac{\partial^2}{\partial v^2}\right ] p(x,v,N ; t) \nl && + \sum_{N' = 0,1} R_{N N'}(x) p(x, v, N'; t) ,
\la{eq:stocheq}
\eea
where $ D $ is a diffusion constant and  
\be
R_{00}(x) = -\gu (x), \quad R_{01}(x) = \gd(x ), \quad R_{10}(x) = \gu(x), \quad R_{11}(x) = -\gd(x).
\la{Rmatrix}
\ee

We again decompose $\gamma = \gamma_{\rm diss} + \gamma_{\rm load}$ and assume a fluctuation-dissipation relation for the dissipative part only
\be
D =  \frac{\gamma_{\rm diss}}{m \beta_O} ,
\la{eq:FD}
\ee
where the oscillator of $m$ is taken to be immersed in a bath at inverse temperature $\beta_O$.  Note that in the limit of large $m$ the dot-oscillator's thermal fluctuations become negligible and the deterministic model of \Sec{sec:deterministic} is a good approximation.  Following the reasoning in \Sec{sec:ST}, we write the second law (in $k_B = 1$ units) for the oscillator as
\be
{\cal S} (t) = -\sum_{N = 0,1} \int dx \,dv \, p(x, v, N; t) \ln p(x, v, N; t) .
\la{eq:entropy_dot_p}
\ee
The right-hand side of the evolution \Eq{eq:stocheq} can be decomposed into the contributions corresponding to the various reservoirs:
\be
\frac{\partial \vv p}{\partial t} =  \La \vv p = \left( \La_O  +  \La_L + \La_R + \La_{\rm work} \right) \vv p . 
\la{eq:stocheqdec}
\ee 
The first generator describes the evolution of the dot under the influence of oscillator's bath only and is given by the following differential operator:
\be
\La_O = - v\frac{\partial}{\partial x} +\frac{\partial}{\partial v} \left(\Omega_0^2 x + a N x+\gamma_{\rm diss} v\right) + D \frac{\partial^2}{\partial v^2} ,
\la{eq:LO}
\ee 
which possesses the stationary distribution in the Gibbs form
\be
\bar p_O(x,v,N) = Z^{-1} e^{-\beta_O E(x, v, N)} .
\la{eq:LOst}
\ee 
Note that this Gibbs distribution is a stationary one for the full evolution \Eq{eq:stocheq} under the thermodynamic equilibrium conditions $\beta_O = \beta_L = \beta_R $,  $\tilde \mu_L = \tilde \mu_R$,  $\gamma_{\rm load} = 0$.
\par
The next two generators describe the influence of each of the electrodes and can be written as $2 \times 2$ matrix-valued functions of $x$
\be
\La_X = \left( \begin{array}{r r}
- \gu^X(x) & \gd^X(x) \\
\gu^X(x) & - \gd^X(x) 
\end{array} \right)
\la{eq:LX}
\ee
for $X= L,R$, with the stationary solutions
\be
\bar p_X(x,v,0) = k(x,v) \frac{\gd^X(x)}{\gd^X(x) + \gu^X(x)}  \quad \hbox{and} \quad
\bar p_X(x,v,1) = k(x,v) \frac{\gu^X(x)}{\gd^X(x) + \gu^X(x)} ,
\la{eq:LXst}
\ee
where $k(x,v)$ is an arbitrary probability distribution. 

The KMS condition of \Eq{eq:KMS} implies a stationary distribution of the form
\be
\bar p_X(x,v,N)  = Z^{-1}_X(x,v) e^{-\beta_X (\epsilon +a x -\mu_X) N} .
\la{eq:LXst1}
\ee
The last dissipator
\be
\La_{\rm work} = \gamma_{\rm load} \frac{\partial}{\partial v} v
\la{eq:LW}
\ee
describes the damping corresponding to energy transfer to the external load (work reservoir). However, this dissipator does not satisfy the condition of \Eq{eq:workdiss}, but rather
\be
\langle \La^\ast_{\rm work} \ln \vv p \rangle_{\vv p} = \gamma_{\rm load}
\la{eq:LWentropy}
\ee
for any probability distribution $\vv p$.  This implies the the loading force contributes negatively to the entropy production rate, but this is a spurious effect due to the fact that we have not tried to describe the dynamics associated the rectification needed for the dot's oscillation to ``raise'' the external load (see the discussion about the one-way valves at the end of \Sec{sec:active}).  If we ignore the degrees of freedom involved in this rectification, the pure damping of \Eq{eq:LW} appears to shrink the phase-space volume for the system.

The problem of the rectification needed for an oscillating engine to raise its load calls for a much deeper investigation, both in general thermodynamic terms and in light of specific implementations of practical importance.  The well developed theory of particle transport (pumping) by ratchets driven by an external periodic modulation (see \cite{Hanggi} and references therein) is evidently relevant here.  In \cite{LEC} we showed that a self-oscillating electrical double layer (EDL) generates an electromotive force (emf), despite the absence of a coherent and time-varying magnetic flux.  This emf can pump a direct current (DC).  In that case, rectification results from the asymmetry between the large (ideally constant) electric field inside the EDL and the small (ideally zero) field outside it \cite{LEC}.  This may be relevant to a wide variety of natural and technological DC generators, but in this article we will not pursue further the detailed dynamics of rectification in oscillating engines.

The entropy balance for our model of the oscillating engine can be written in the form of the second law for open systems:
\be
\dot {\cal S} - \sum_{Y = O, L, R} \beta_Y j^h_Y +\gamma_{\rm load} \geq 0 .
\la{eq:IIlawstoch}
\ee
However, as explained above the entropy outflow represented by $\gamma_{\rm load}$ in \Eq{eq:IIlawstoch} is spurious, and may in any case be neglected compared to the other terms.  The heat currents are defined as
\be
j^h_O = -(\beta_O)^{-1} \langle \La_O^\ast\ln \bar{\vv p}_O \rangle_{\vv p} = -\gamma_{\rm diss} \langle v^2 \rangle_{\vv p} , 
\la{eq:heatO}
\ee
and
\be
j^h_X = - (\beta_X)^{-1} \langle \La_X^\ast\ln \bar{\vv p}_X \rangle_{\vv p} =\langle (\epsilon + a x - \mu_X) \La_X^\ast N \rangle_{\vv p} .
\la{eq:heatX}
\ee
for $X= L,R$.  Introducing the internal energy of the dot-piston with its time derivative
\be
{\cal U} = \langle U(x, v, N) \rangle_{\vv p} , \quad \dot {\cal U} = \langle \La^\ast U \rangle_{\vv p} .
\la{eq:Ienergy}
\ee
Using Eqs.\ \eqref{eq:heatO} and \eqref{eq:heatX}, one can express the energy balance in the form of first law of thermodynamics
\be
\dot {\cal U} = \sum_{Y= O, L, R} j^h_Y + \sum_{X= L,R} \mu_X \dot \Na_X - {\cal P}
\la{eq:Ilawstoch}
\ee
with the output power
\be
{\cal P} = \gamma_{\rm load} \left( \left\langle v^2 \right\rangle - \beta_O^{-1} \right) .
\la{eq:powerstoch}
\ee
Note that positive power is generated when the energy of the self-oscillating piston exceeds its thermal energy.


\subsection{Stochastic model of rotating engine}
\la{sec:stochastic-rot}

\begin{figure} [t]
\centering
	\subfigure[]{\includegraphics[height=0.32 \textwidth]{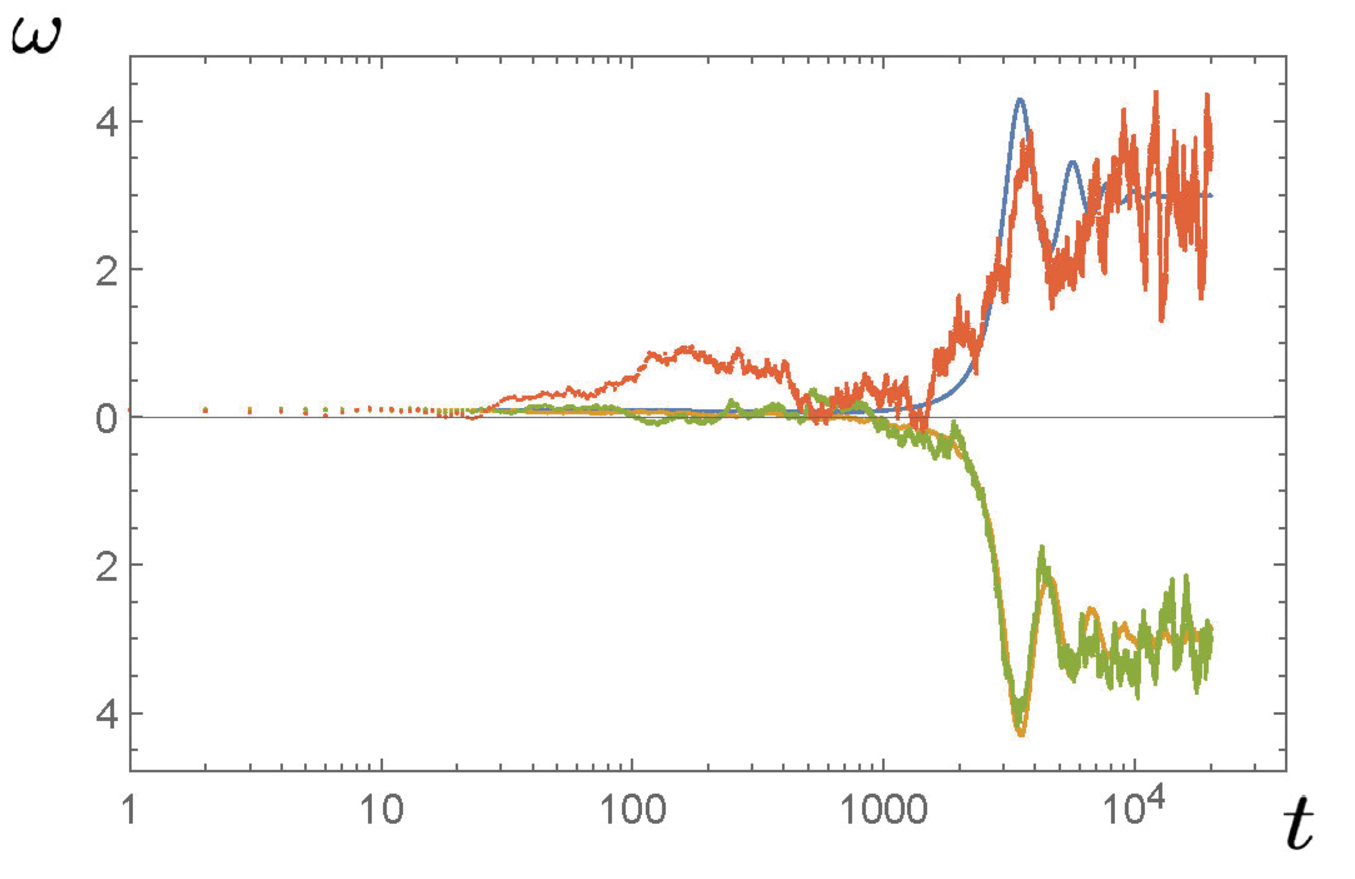}}
	\subfigure[]{\includegraphics[height=0.32 \textwidth]{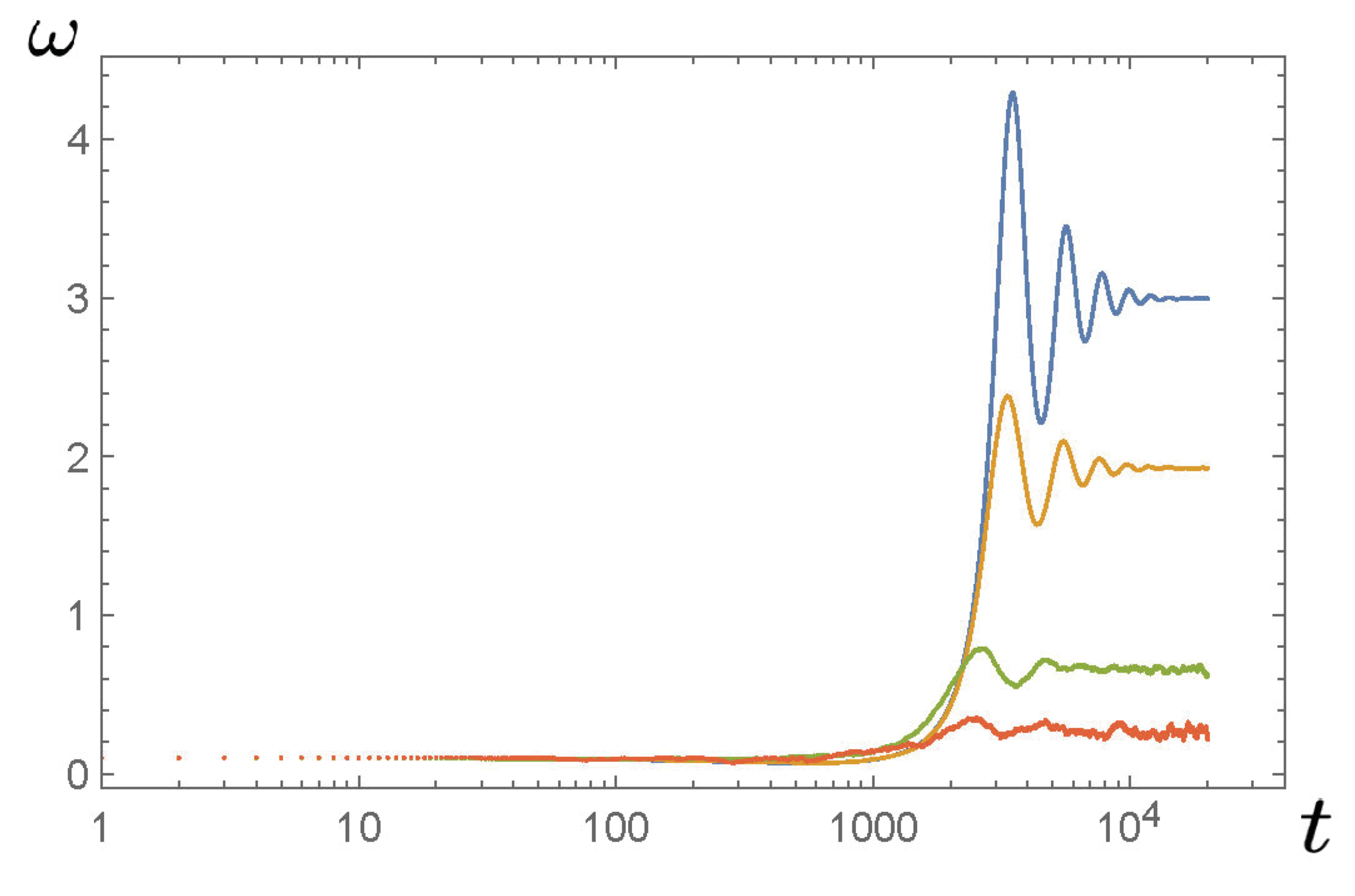}}
\caption{\small Plots of $\omega(t)$ for the model of the Quincke rotating engine described by \Eq{eq:langevin_tau}.  Time $t$ is measured in units of $1/\gamma$.  The parameters are $\mu = 10$, $\Gamma = 1$, $J = 1$, and $\tau_{\rm load} = 0.01$.  The colors represent different amplitudes for the white noise term $\tau_L(t)$, with blue corresponding to 0, yellow to 0.1, green to 0.5, and dark orange to 1.  The initial conditions are $\omega(0) = 0.1 $ and $z(0) = 0$.  Plot (a) shows single realizations of the stochastic trajectory, while plot (b) shows the averages over 500 realizations.\la{fig:rotor-noise}}
\end{figure}

One can easily construct the Fokker-Planck equation describing time evolution of the probability distribution $p(\omega, z, \bar z; t)$, for the rotor coupled to oscillator, generated by the stochastic extension of \Eq{eq:dynsystem1}. One can repeat all the derivations of thermodynamic relations, similarly to the case of electron shuttle. In particular, here the problem of spurious entropy outflow does not occur as the loading is simply described by the conservative gravitational torque. Indeed, the corresponding
\be
\La_{\rm load} = -\tau_{\rm load} \, \frac{\partial}{\partial\omega}
\ee
preserves entropy.

To avoid repetition, we present only numerical examples of trajectories which are the solutions  of the Langevin equation
\be
\left\{ \begin{array}{l}
\displaystyle \dot \omega = - \gamma\omega +\frac{\mu}{2i} (z - \bar z) - \tau_{\rm load} + \tau_L (t) \\
\displaystyle \dot z = (i \omega - \Gamma ) z + J \phantom{+ \frac 0 1} \end{array} \right. ,
\la{eq:langevin}
\ee 
with the stochastic white noise $\tau_L(t)$ satisfying fluctuation-dissipation relation 
\be
\langle \tau_L (t) \rangle = 0 \quad \hbox{and} \quad \langle \tau_L (t) \tau_L (s) \rangle = \frac{2 \gamma}{\beta I} \delta(t-s) ,
\la{eq:langevin_tau}
\ee
where $\beta$ is the bath's inverse temperature.  Here, we do not add noise to the equation for the complex variable $z$ describing the shape of surface charge distribution, because we have neglected the diffusion of charge along the surface of the cylinder.

To compute the magnitude of the noise in \Eq{eq:langevin_tau} one takes the asymptotic value of the average squared angular velocity for the free rotor (i.e. $\mu = \tau_{\rm load}=0$), which is
\be
\left\langle \omega^2 \right\rangle =  \int_0^{\infty}dt \int_0^{\infty}ds \, e^{-\gamma(t+s)} \langle\tau_L (t) \tau_L (s)\rangle ,
\la{eq:langevin_tau1}
\ee
and inserts this into the equilibrium relation for the rotor kinetic energy
\be
\frac 1 2 I \left\langle \omega^2 \right\rangle = \frac 1 2 \beta^{-1}. 
\la{eq:langevin_tau2}
\ee

Figure \ref{fig:rotor-noise} illustrates the behavior of the model of \Eq{eq:langevin_tau} for various choices the amplitude of the noise term $\tau_L (t)$ (corresponding to varying the temperature of the medium).  If the trajectory settles to a positive value of $\omega$, this indicates a steady operation of the engine with positive power output $\omega \cdot \tau_{\rm load}$.  It is interesting to note that in \Fig{fig:rotor-noise}(a) the trajectory with the largest noise (shown in dark orange) appears to settle to a steady positive work output similar to that of the noiseless engine, while less noisy realizations (shown in yellow and green curves) do not yield positive power.  On the other hand, \Fig{fig:rotor-noise}(b) shows that when the trajectories are averaged over many realizations, increasing the amplitude of the noise reduces the engine's steady power output, as one might expect.  The dynamics of the noisy Quincke rotor deserves much more detailed investigation, especially because it may have significant applications in biophysics and nanotechnology.


\subsection{The puzzle of stationary states}
\la{sec:stationary}

The main lesson that we have sought to transmit in this article is that any effective description of an engine that omits the dynamics of the tool and its feedback with the working substance is leaving out something important, without which a thermodynamically complete understanding of the engine's operation is not possible.  The dynamics of the tool can be described as a self-oscillation or a self-rotation, corresponding to a limit cycle in  phase space.  When thermal noise is included, we see that the trajectories (i.e., solutions to the generalized Langevin equation) for the tool degree of freedom fluctuate about the curve corresponding to that limit cycle.

On the other hand, we have seen in this section that the full stochastic model that includes the tool degree of freedom in the Fokker-Planck (master) \Eq{eq:stocheq} possesses a unique stationary state $\bar p(x, v, N)$.  In \Sec{sec:stochastic-osc} we were able to define the engine's output power by using a phenomenological picture of pure ``damping'' by the external load.  This $\bar p$ can therefore be interpreted as a non-equilibrium steady state (NESS) for the running engine, from which work can be extracted because the average kinetic energy exceeds its thermal equilibrium value (\Eq{eq:powerstoch}).  This raises the question of how to see the active dynamics of the tool in the stationary probability distribution $\bar p$.

We expect that evidence of the cyclic deterministic dynamics (e.g., the self-oscillation of the grain in \Fig{fig:shuttle}) must lie in the background of the stochastic system that we set up by adding the thermal noise, and that this evidence can be found using the temporal correlation functions for the dynamical variables.  For example, for a system in which the time-evolution is generated by $\La$ (see \Eq{eq:stocheqdec}), an auto-correlation function of the form
\be
C_x(t) =  \left\langle x \, e^{\La^\ast t}x \right\rangle - \langle x \rangle^2 ,
\la{eq:corr}
\ee
computed for the NESS given by $\bar p$, should reveal piston's self-oscillations, with frequency close to $\Omega_0$ and thermally damped by the phase diffusion introduced by the noise.  This can be formulated in terms of the topological distinction between a random walk around a fixed point (i.e., the Brownian motion of a passive particle) and a noisy limit-cycle trajectory in phase-space (corresponding to the active dynamics of a steadily running engine).  This important question calls for a much more careful mathematical formulation and investigation.

We do not have, at the moment, adequate mathematical tools to study and characterize the statistical properties of active systems like the engines we have considered in \Sec{sec:dynamics}, combining deterministic feedbacks leading to active forces with stochastic fluctuations.  A further complication is introduced by the fact that the fluctuations may, in some cases, make the active dynamics intermittent by pushing the dynamical system through Hopf bifurcations; see, e.g., \cite{on-off, Hopf}.   Such intermittency may be relevant not just in biophysics and microengineering, but also in turbulence \cite{Frisch}, and even in very large active systems such as geophysics \cite{Chandler} and macroeconomics \cite{SO, macro}.  The development of an adequate statistical treatment of such non-Gaussian systems remains a great intellectual challenge.


\section{Discussion and outlook}
\la{sec:discussion}

In this article we have sketched what we consider to be a physically realistic approach to the dynamics of engines.  We showed that the laws of thermodynamics imply that any open system capable of generating net work without ``running away'' in configuration space must exhibit a cyclical time dependence of its macroscopic state, which corresponds to the engine's working substance executing what textbooks identify as that engine's ``thermodynamic cycle''.  This is incompatible with the conceptualizations of active systems as ``dissipative structures'' or ``non-equilibrium steady states'' that have dominated the theoretical literature on non-equilibrium thermodynamics.

The engine's work output is seen in that the working substance exerts an {\it active force} on a macroscopic and mechanical degree of freedom endowed with inertia, which we generically call the ``tool''.  This active force is not given by the gradient of any mechanical or thermodynamic potential.  Instead, it arises from a positive feedback between the state of the tool and the coupling of the working substance to the external baths whose disequilibrium powers the engine.  Such a feedback is necessarily irreversible, and therefore can appear only in the presence of dissipation.  While this feedback operates, the active force prevents the tool from finding any equilibrium.

We have illustrated our approach with simple examples of both oscillating and rotating engines.  The relation between the two is not as obvious as one might expect, and the mathematical treatment needed for each is somewhat different, even though their thermodynamics is very similar.  One clear difference between the two kinds of engines is that an oscillating engine requires an additional process of rectification to raise a load, while a rotating engine can raise the load directly.  Our discussion of the self-oscillating ``putt-putt'',  Franklin bells, and electron shuttle engines is based on previously published work by the authors \cite{AGJ, LEC} and others \cite{stochastic-shuttle, Strasberg}, but the formulation of our model of the Quincke rotor in \Sec{sec:Quincke} is largely original and its applications bear further study.

An important part of our approach, in which it differs from previous work about autonomous engines in quantum thermodynamics, was to distinguish between the {\it tool} that the active force acts upon and the {\it load}, external to the engine, that extracts work (zero-entropy energy) from the tool.  We have argued that this clarifies the distinction between the heat and the work outputs by the engine, facilitating the formulation of thermodynamically complete equations of motion that incorporate noise into a well posed extension of the Fokker-Planck and Langevin equations.  As stressed in \cite{Torun}, finding a consistent operational definition of work in a microscopic model of an engine is a theoretical question that bears some similarity with the problem of quantum measurement.  In both cases, the cleanest solution is based on introducing a classical measurement device, which in our model for the engine is the external load.  Only when the engine's {\it tool} is also macroscopic (which is not the case in mesoscopic engines, such as those seen in biological cells), can its fluctuations be neglected.

This formulation of engine dynamics, using the Markovian master equation for open systems, allowed us finally to describe the engine's steady operation in terms of a stationary state $\bar p(x, v, N)$ for the combination of the working substance {\it and} the tool, which are coupled to an external environment that includes a load.  We have argued that this incorporates thermal noise in a thermodynamical consistent way.  However, the mathematical characterization of such stationary but active distributions $\bar p$ requires much further development, as underlined in \Sec{sec:stationary}.

We hope that the present article helps to clarify how this approach differs from other theoretical treatments to active systems in statistical physics, and why we believe that it is more physically realistic.  As we underlined in \Sec{sec:stationary}, however, the existing mathematical tools used by statistical physicists are not well suited for capturing the dynamics of active systems.  We believe that the difficulties with which statistical physics has repeatedly met in describing ``complex systems'' may have as much to do with complexity {\it per se} as with the lack of a realistic description of active forces, which are ubiquitous both in nature and in technology.  Much work is needed, both in the mathematical and the physical sides of this problem, but we are hopeful that it is now possible to discern a way forward in this field.

One important question that should be pursued within this framework is the determination of the efficiency limits on engines at maximum power, a subject which has thus far been studied only for heat engines and without formulating equations of motion (see \cite{Ouerdane} and references therein).  Chemical engines (which are of enormous importance in biophysics) are not Carnot-bounded, so that their efficiencies can approach arbitrarily close to unity if the corresponding cycle is sufficiently slow and therefore close to being reversible.  But the need for irreversible dynamics in the equations of motion that we studied in \Sec{sec:dynamics} implies that any autonomous engine that can generate finite power must consume free energy.  As we underlined in \Sec{sec:RE} such irreversible dynamics cannot be understood using the maximum or minimum entropy production principles that have been applied successfully to passive or non-cyclical processes.

A related open question concerns the theory of the precision and stability of clocks, and how their robustness to noise is related to their consumption of free energy.  This problem has attracted considerable attention both in engineering (see, e.g., \cite{Groszkowski, Harrison}) and biophysics (see, e.g., \cite{uncertainty, oscillation-cost}), but we believe that greater theoretical clarity requires a realistic consideration of the dynamics of clocks considered as autonomous engines.

Another important question that we believe should be studied in this framework is the thermodynamic description of non-linear friction and damping terms in many macroscopic systems of interest.  Only a damping linear in the velocity ($- \gamma \vv v$) can be obtained directly from a thermodynamically acceptable Langevin equation.  We suggest that the more complicated forms of damping seen experimentally should be explained by loading forces associated with the extraction of work, which is dissipated into heat only at a later stage.  Note that such non-linear damping is needed to obtain a limit cycle in mathematical models of self-oscillation \cite{Strogatz, SO}. Recent work by two of the authors on the triboelectric effect and its association with non-linear dry friction suggests that the tools of quantum thermodynamics may be useful in this context \cite{tribo}.

Given the challenges that we have explored here to describing simple engines with the current tools of mathematical physics, we expect that the study of ``active matter'' (such as swarms of bacteria or flocks of birds) in statistical mechanics and soft condensed-matter physics may benefit from new approaches.  ``Active matter'' is usually defined as being composed of elements (such as an individual bacterium) that can consume and dissipate energy, in the process executing systematic movement \cite{Ramaswamy}.  The literature on active matter in statistical mechanics and soft condensed matter theory accounts for the energy consumed by those elements from an underlying thermodynamic disequilibrium (see \cite{Cates} for recent work in this area), but it does not treat the dynamics of the work extraction in a physically realistic way, as we stressed in \cite{LEC}.  From our point of view, the great scientific challenge before us is to understand the dynamics of the active forces that give bacteria and birds their {\it \'elan vital}.


\begin{acknowledgments} We thank Elizabeth von Hauff for extensive discussions on various issues treated in this work.  RA thanks Philipp Strasberg for stimulating discussions regarding the methods of \cite{stochastic-shuttle, shuttle-rotor}.  AJ thanks Baran \c{C}imen for suggesting consideration of the Quincke rotor and for other encouraging conversations.  RA was supported by the International Research Agendas Programme (IRAP) of the Foundation for Polish Science (FNP), with structural funds from the European Union (EU). DG-K was supported by the Gordon and Betty Moore Foundation as a Physics of Living Systems Fellow (grant no.\ GBMF4513).  AJ was supported by the Polish National Agency for Academic Exchange (NAWA)'s Ulam Programme (project no.\ PPN/ULM/2019/1/00284). \end{acknowledgments}



\begin{thebibliography}{99}


\bibitem{Schroeder}
	D.~V.~Schroeder,
	{\it An Introduction to Thermal Physics}
	(San Francisco: Addison Wesley Longman, 2000), sec.\ 4.1
	
\bibitem{QT}
	R.~Alicki and R.~Kosloff,
	``Introduction to quantum thermodynamics: History and prospects'',
	in {\it Thermodynamics in the Quantum Regime}, eds.\ F.~Binder et al.,
	(Cham: Springer, 2019), \href{https://doi.org/10.1007/978-3-319-99046-0_1}{pp.\ 1--33}
	[arXiv:1801.08314 [quant-ph]]
	
\bibitem{current}
	T.~N.~Todorov, D.~Dundas, J.-T.~L\"u, M.~Brandbyge, and P.~Hedeg\r{a}rd,
	``Current-induced forces: a simple derivation'',
	Eur.\ J.\ Phys.\ \href{https://doi.org/10.1088/0143-0807/35/6/065004}{{\bf 35}, 065004} (2014)
	[arXiv:1405.4381 [cond-mat.mes-hall]]
	
\bibitem{AQM1}
	R.~Bustos-Mar\'un, G.~Refael, and F.~von Oppen,
	``Adiabatic quantum motors'',
	Phys.\ Rev.\ Lett.\ \href{https://doi.org/10.1103/PhysRevLett.111.060802}{{\bf 111}, 060802} (2013)
	[arXiv:1304.4969 [cond-mat.mes-hall]]
	
\bibitem{AQM2}
	A.~Bruch, S.~V.~Kusminskiy, G.~Refael, and F.~von Oppen,
	``Interacting adiabatic quantum motor'',
	Phys.\ Rev.\ B \href{https://doi.org/10.1103/PhysRevB.97.195411}{{\bf 97}, 195411} (2018)
	[arXiv:1712.04952 [cond-mat.mes-hall]]
	
\bibitem{Kubo}
	R.~Kubo,
	``The fluctuation-dissipation theorem'',
	Rep.\ Prog.\ Phys.\ \href{https://doi.org/10.1088/0034-4885/29/1/306}{{\bf 29}, 255} (1966)
	
\bibitem{Onsager1}
	L.~Onsager,
	``Reciprocal Relations in Irreversible Processes I'',
	Phys.\ Rev.\ \href{http://doi.org/10.1103/PhysRev.37.405}{{\bf 37}, 405} (1931)
	
\bibitem{Onsager2}
	L.~Onsager,
	``Reciprocal Relations in Irreversible Processes II'',
	Phys. Rev.\ \href{http://doi.org/10.1103/PhysRev.38.2265}{{\bf 38}, 2265} (1931)
	
\bibitem{MEPP}
	L.~M.~Martyushev and V.~D.~Seleznev,
	``Maximum entropy production principle in physics, chemistry and biology'',
	Phys.\ Rep.\ \href{https://doi.org/10.1016/j.physrep.2005.12.001}{{\bf 426}, 1} (2006)
	
\bibitem{Landauer1}
	R.~Landauer,
	``Stability and entropy production in electrical circuits'',
	J.\ Stat.\ Phys.\ \href{https://doi.org/10.1007/BF01012596}{{\bf 13}, 1} (1975)
	
\bibitem{Landauer2}
	R.~Landauer,
	``Inadequacy of entropy and entropy derivatives in characterizing the steady state'',
	Phys.\ Rev.\ A \href{https://doi.org/10.1103/PhysRevA.12.636}{{\bf 12}, 636} (1975)

\bibitem{Prigogine}
	G.~Nicolis and I.~Prigogine,
	{\it Self-Organization in Nonequilibrium Systems}
	(New York: Wiley, 1977)
	
\bibitem{Haken}
	H.~Haken,
	{\it Synergetics: An Introduction},
	(Berlin: Springer-Verlag, 1977)
		
\bibitem{Anderson}
	P.~W.~Anderson and D.~L.~Stein,
	``Broken Symmetry, Emergent Properties, Dissipative Structures, Life: Are They Related?"
	in {\it Self-Organizing Systems: The Emergence of Order}, ed.\ F.~E.~Yates, 
	(New York: Plenum Press, 1987), \href{https://doi.org/10.1007/978-1-4613-0883-6_24}{pp.\ 445--57}
	
\bibitem{LEC}
	R.~Alicki, D.~Gelbwaser-Klimovsky, and A.~Jenkins,
	``Leaking elastic capacitor as model for active matter'',
	Phys.\ Rev.\ E \href{https://doi.org/10.1103/PhysRevE.103.052131}{{\bf 103}, 052131} (2021)
	[arXiv:2010.05534 [physics.class-ph]]
	
\bibitem{AVK}
	A.~A.~Andronov, A.~A.~Vitt, and S.~\`E.~Kha\u{\i}kin,
	{\it Theory of Oscillators}, ed.\ W.~Fishwick,
	(Mineola, NY: Dover, 1987 [1966]), p.\ 200
	
\bibitem{Airy}
	G.~B.~Airy,
	``On certain Conditions under which a Perpetual Motion is possible,''
	Trans.\ Cambridge Phil.\ Soc.\ \href{http://books.google.com/books?id=oxpYAAAAYAAJ&pg=PA369}{{\bf 3}, 369} (1830)
	
\bibitem{ToS}
	J.~W.~Strutt (Lord Rayleigh),
	{\it The Theory of Sound}, 2nd ed., vols.\ I and II,
	(New York: Dover, 1945 [1896]),
	secs.\ 66a, 68a--68b, 103a--103b, 322a--322k, 397, appendix to ch.\ V
	
\bibitem{Strogatz}
	S.~H.~Strogatz,
	{\it Nonlinear Dynamics and Chaos}, 2nd ed.,
	(Boulder: Westview Press, 2014)
	
\bibitem{Kirillov}
	O.~N.~Kirillov,
	{\it Nonconservative Stability Problems of Modern Physics}, 2nd ed.,
	(Berlin: De Gruyter, 2021)
	
\bibitem{LeC1}
	P.~Le Corbeiller,
	``The non-linear theory of the maintenance of oscillations,''
	J.\ Inst.\ Electr.\ Engrs.\ {\bf 79}, 361 (1936),
	reprinted in Proc.\ Inst.\ Electr.\ Engrs.\ \href{https://doi.org/10.1049/PWS.1936.0030}{{\bf 11}, 292} (1936)
	
\bibitem{LeC2}
	P.~Le Corbeiller,
	``Theory of prime movers'',
	in {\it Non-Linear Mechanics}, eds.\ K.~O.~Friedrichs, P.~Le Corbeiller, N.~Levinson and J.~J.~Stoker,
	(Providence: Brown U., 1943), pp.\ 2.1--2.18
	
\bibitem{SO}
	A.~Jenkins,
	``Self-oscillation'',
	Phys.\ Rep.\ \href{https://doi.org/10.1016/j.physrep.2012.10.007}{{\bf 525}, 167} (2013)
	[arXiv:1109.6640 [physics.class-ph]]
	
\bibitem{Lugt}
	H.~J. Lugt,
	``Autorotation'',
	Ann.\ Rev.\ Fluid Mech.\ \href{https://doi.org/10.1146/annurev.fl.15.010183.001011}{{\bf 15}, 123} (1983)
	
\bibitem{rotors}
	S.~Seah, S.~Nimmrichter, and V.~Scarani,
	``Work production of quantum rotor engines'',
	New J.\ Phys.\ \href{https://doi.org/10.1088/1367-2630/aab704}{{\bf 20}, 043045} (2018)
	[arXiv:1801.02820 [quant-ph]]
	
\bibitem{Bob-lift}
	R.~L.~Jaffe and W.~Taylor,
	{\it The Physics of Energy}
	(Cambridge: Cambridge University Press, 2018), chs.\ 29--30
	
\bibitem{SA-lift}
	E.~Regis,
	 ``The Enigma of Aerodynamic Lift'',
	  Sci.\ Am.\ \href{10.1038/scientificamerican0220-44}{{\bf 322}(2), 44} (2020)
	
\bibitem{Nakanishi}
	H.~Nakanishi, S.~Fujiwara, K.~Takayama, I.~Kawayama, H.~Murakami, and M.~Tonouchi,
	``Imaging of a Polycrystalline Silicon Solar Cell Using a Laser Terahertz Emission Microscope'',
	Appl.\ Phys.\ Express \href{https://doi.org/10.1143/APEX.5.112301}{{\bf 5}, 112301} (2012)

\bibitem{Guzelturk}
	B.~Guzelturk et al.,
	``Terahertz Emission from Hybrid Perovskites Driven by Ultrafast Charge Separation and Strong Electron--Phonon Coupling'',
	Adv.\ Mater.\ \href{https://doi.org/10.1002/adma.201704737}{{\bf 30}, 1704737} (2018)
	
\bibitem{solarcells}
	R.~Alicki, D.~Gelbwaser-Klimovsky, and K.~Szczygielski,
	``Solar cell as a self-oscillating heat engine'',
	J.\ Phys.\ A: Math.\ Theor.\
	\href{http://dx.doi.org/10.1088/1751-8113/49/1/015002}{{\bf 49}, 015002} (2016)
	[arXiv:1501.00701 [cond-mat.stat-mech]]
	
\bibitem{thermocells}
	R.~Alicki,
	``Thermoelectric generators as self-oscillating heat engines'',
	J.\ Phys.\ A: Math.\ Theor.\
	\href{http://dx.doi.org/10.1088/1751-8113/49/8/085001}{{\bf 49}, 085001} (2016)
	[arXiv:1506.00094 [quant-ph]]
	
\bibitem{fuelcells}
	R.~Alicki,
	``Unified Quantum Model of Work Generation in Thermoelectric Generators, Solar and Fuel Cells'',
	Entropy \href{http://dx.doi.org/10.3390/e18060210}{{\bf 18}, 210} (2016)
	
\bibitem{AGJ}
	R.~Alicki, D.~Gelbwaser-Klimovsky, and A.~Jenkins,
	``A thermodynamic cycle for the solar cell'',
	Ann.\ Phys.\ (NY)
	\href{https://doi.org/10.1016/j.aop.2017.01.003}{{\bf 378}, 71} (2017)
	[arXiv:1606.03819 [cond-mat.stat-mech]]
	
\bibitem{Torun}
	R.~Alicki,
	``From the GKLS equation to the theory of solar and fuel cells'',
	Open Syst.\ Inf.\ Dyn.\ \href{https://doi.org/10.1142/S1230161217400078}{{\bf 24}, 1740007} (2017)
	[arXiv:1706.10257 [quant-ph]]
	
\bibitem{solar-dyn}
	R.~Alicki, D.~Gelbwaser-Klimovsky, A.~Jenkins, and E.~von Hauff,
	``A dynamic picture of energy conversion in photovoltaic devices'',
	arXiv:1901.10873 [physics.app-ph]
	
\bibitem{battery}
	R.~Alicki, D.~Gelbwaser-Klimovsky, A.~Jenkins, and E.~von Hauff,
	``A dynamical theory of the battery's electromotive force'',
	Phys.\ Chem.\ Chem.\ Phys.\ \href{http://doi.org/10.1039/D1CP00196E}{{\bf 23}, 9428} (2021)
	[arXiv:2010.16400 [physics.chem-ph]]
	
\bibitem{Goupil}
	C.~Goupil, H.~Ouerdane, E.~Herbert, G.~Benenti, Y.~D'Angelo, and Ph.~Lecoeur,
	``Closed-loop approach to thermodynamics'',
	Phys.\ Rev.\ E {\bf 94}, 032136 (2016)
	[arXiv:1606.03387 [cond-mat.stat-mech]]
	
\bibitem{stochastic-shuttle}
	C.~W.~W\"achtler, P.~Strasberg, S.~H.~L.~Klapp, G.~Schaller, and C.~Jarzynski,
	``Stochastic thermodynamics of self-oscillations: the electron shuttle'',
	New J.\ Phys.\ \href{https://doi.org/10.1088/1367-2630/ab2727}{{\bf 21}, 073009} (2019)
	[arXiv:1902.08174 [cond-mat.stat-mech]]
	
\bibitem{shuttle-rotor}
	C.~W.~W\"achtler, P.~Strasberg, and G.~Schaller,
	``Proposal of a Realistic Stochastic Rotor Engine Based on Electron Shuttling'',
	Phys.\ Rev.\ Applied {\bf 12}, 024001 (2019)
	[arXiv:1903.07500 [cond-mat.stat-mech]]
	
\bibitem{Strasberg}
	P.~Strasberg, C.~W.~W\"achtler, and G.~Schaller,
	``Autonomous implementation of thermodynamic cycles at the nanoscale'',
	Phys.\ Rev.\ Lett.\ \href{https://doi.org/10.1103/PhysRevLett.126.180605}{{\bf 126}, 180605} (2021)
	[arXiv:2101.05027 [quant-ph]]


\bibitem{Ouerdane}
	H.~Ouerdane, Y.~Apertet, C.~Goupil, and P.~Lecoeur,
	``Continuity and boundary conditions in thermodynamics: From Carnot's efficiency to efficiencies at maximum power'',
	Eur.\ Phys.\ J.-Spec.\ Top.\ \href{http://dx.doi.org/10.1140/epjst/e2015-02431-x}{{\bf 224}, 839} (2015)
	[arXiv:1411.4230 [physics.hist-ph]]
	
\bibitem{dissipation-induced}
	C.~D.~D\'iaz-Mar\'in and A.~Jenkins,
	``A physical approach to dissipation-induced instabilities'',
	arXiv:1806.01527 [physics.class-ph]
	

\bibitem{Jaynes}
	E.~T.~Jaynes,
	``The muscle as an engine'', unpublished manuscript (1983)
	\url{https://bayes.wustl.edu/etj/articles/muscle.pdf} (accessed 26 Jul.\ 2021)

\bibitem{Eddington1}
	A.~S.~Eddington,
	``The Pulsations of a Gaseous Star and the Problem of the Cepheid Variables II'',
	Mon.\ Not.\ R.\ Astron.\ Soc.\
	\href{http://dx.doi.org/10.1093/mnras/79.3.177}{{\bf 79}, 177--189} (1919)
	
\bibitem{Eddington2}
	A.~S.~Eddington,
	{\it The Internal Constitution of the Stars}
	(Cambridge: Cambridge University Press, 1988 [1926]), \S 134--138

\bibitem{Rayleigh1}
	J.~W.~Strutt (Lord Rayleigh),
	``The Explanation of Certain Acoustical Phenomena'',
	Nature
	\href{http://dx.doi.org/10.1038/018319a0}{{\bf 18}, 319--321} (1878)
	
\bibitem{Rayleigh2}
	J.~W.~Strutt (Lord Rayleigh),
	{\it The Theory of Sound}, 2nd ed., vol.\ II,
	(New York: Dover, 1945 [1896]), \S 322g
	
\bibitem{Rochas}
	A.~Beau de Rochas,
	{\it Nouvelles recherches sur les conditions pratiques de plus grande utilisation de la chaleur et, en g\'en\'eral, de la force motrice}
	(Paris: E.~Lacroix, 1862)
	
\bibitem{motors}
See, e.g.,
	Y.~A.~\c{C}engel and M.~A.~Boles,
	{\it Thermodynamics: An Engineering Approach}, 8th ed.
	(New York: McGraw-Hill, 2011) 

\bibitem{Stirling}
	G.~Walker,
	{\it Stirling Engines}
	(Oxford: Oxford University Press, 1980
	
	
\bibitem{Rayleigh-vdP}
	J.~W.~Strutt (Lord Rayleigh),
	``On maintained vibrations,''
	Philos.\ Mag.\ (ser.\ 5) \href{http://dx.doi.org/10.1080/14786448308627342}
	{{\bf 15}, 229-235} (1883)

\bibitem{vdP1}
	B.~van der Pol,
	``On `relaxation-oscillations,' \!\!''
	Philos.\ Mag.\ (ser.\ 7) \href{https://doi.org/10.1080/14786442608564127}{{\bf 2}, 978} (1926)
	
\bibitem{vdP2}
	B.~van der Pol,
	``Forced Oscillations in a Circuit with non-linear Resistance. (Reception with reactive Triode.)'',
	in {\it Selected Papers on Mathematical Trends in Control Theory},
	eds.\ R.~Bellmann and R.~Kalaba, (New York: Dover, 1964), pp.\ 124--140.
This is a reprinting of
	Philos.\ Mag.\ (ser.\ 7) \href{https://doi.org/10.1080/14786440108564176}{{\bf 3}, 65} (1927)
	
\bibitem{Stuart-Landau}
See Y.~Kuramoto,
	{\it Chemical Oscillations, Waves, and Turbulence}
	(Berlin: Springer-Verlag, 1984), ch.\ 2, and references therein
	
\bibitem{active-brownian}
	W.~Ebeling, F.~Schweitzer, and B.~Tilch,
	``Active Brownian particles with energy depots modeling animal mobility'',
	BioSystems \href{https://doi.org/10.1016/S0303-2647(98)00027-6}{{\bf 49}, 17} (1999)
	
\bibitem{self-propelled}
	C.~Bechinger, R.~Di Leonardo, H.~L\"owen, C.~Reichhardt, G.~Volpe, and G.~Volpe,
	``Active particles in complex and crowded environments'',
	Rev.\ Mod.\ Phys.\ \href{https://doi.org/10.1103/RevModPhys.88.045006}{{\bf 88}, 045006} (2016)
	[arXiv:1602.00081 [cond-mat.soft]]
	
\bibitem{Seifert}
	U.~Seifert,
	``Stochastic thermodynamics, fluctuation theorems and molecular machines'',
	Rep.\ Prog.\ Phys.\
	\href{https://doi.org/10.1088/0034-4885/75/12/126001}{{\bf 75}, 126001} (2012)
	[arXiv:1205.4176 [cond-mat.stat-mech]]
	
\bibitem{Minorsky}
	N.~Minorsky,
	``Self-excited Oscillations in Dynamical Systems Possessing Retarded Action,''
	in {\it Selected Papers on Mathematical Trends in Control Theory},
	eds.\ R.~Bellmann and R.~Kalaba, (New York: Dover, 1964), pp.\ 141--149.
This is a reprinting of
	J.\ Appl.\ Mech.\ {\bf 9}, 65 (1942).

\bibitem{bio-oscillators}
	B.~Nov\'ak and J.~J.~Tyson,
	``Design principles of biochemical oscillators'',
	Nat.\ Rev.\ Mol.\ Cell.\ Biol.\ \href{https://doi.org/10.1038/nrm2530}{{\bf 9}, 981} (2008)
	
\bibitem{Hanggi}
	P.~H\"anggi and F.~Marchesoni,
	``Artificial Brownian motors: Controlling transport on the nanoscale'',
	Rev.\ Mod.\ Phys.\
	\href{https://doi.org/10.1103/RevModPhys.81.387}{{\bf 81}, 387} (2009)
	[arXiv:0807.1283 [cond-mat.stat-mech]]
	
\bibitem{pp1}
	I.~Finnie and R.~L.~Curl,
	``On the functioning of a familiar nonlinear thermodynamic oscillator'',
	Proc.\ Internat.\ Union of Theor.\ and Appl.\ Mech.\
	\href{http://deepblue.lib.umich.edu/handle/2027.42/64001}{{\bf 3}, 486--497} (1963)
	
\bibitem{pp2}
	I.~Finnie and R.~L.~Curl,	
	``Physics in a Toy Boat'',
	Am.\ J.\ Phys.\
	\href{http://dx.doi.org/10.1119/1.1969435}{{\bf 31}, 289--293} (1963)
	
\bibitem{LL}
	L.~D.~Landau and E.~M.~Lifshitz,
	{\it Mechanics}, 3rd ed.,
	(Oxford: Elsevier, 1976), sec.\ 25
	
	
\bibitem{Jefimenko}
	O.~D.~Jefimenko,
	{\it Electrostatic Motors}
	(Star City, WV: Electret Scientific Co., 1973)
	
\bibitem{Pathria}
	R.~K.~Pathria and P.~D.~Beale,
	{\it Statistical Mechanics}, 4th ed.,
	(London: Academic Press, 2022), ch.\ 15
	
\bibitem{shuttle}
	L.~Y.~Gorelik, A.~Isacsson, M.~V.~Voinova, B.~Kasemo, R.~I.~Shekhter, and M.~Jonson,
	``Shuttle Mechanism for Charge Transfer in Coulomb Blockade'',
	Phys.\ Rev.\ Lett.\ \href{https://doi.org/10.1103/PhysRevLett.80.4526}{{\bf 80}, 4526} (1998)
	[arXiv:cond-mat/9711196]
	
\bibitem{Weiler}
	W.~Weiler,
	``Zur Darstellung elektrischer Kraftlinien'',
	Z.\ phys.\ chem.\ Unterricht {\bf 6}, 194 (1893)

\bibitem{Quincke}
	G.~Quincke,
	``Ueber Rotationen im constanten electrischen Felde",
	Ann.\ Phys.\ Chem.\ \href{https://doi.org/10.1002/andp.18962951102}{{\bf 295}, 417} (1896)	
	
\bibitem{Jones}
	T.~B.~Jones,
	{\it Electromechanics of Particles}
	(Cambridge: Cambridge University Press, 1995), \href{https://doi.org/10.1017/CBO9780511574498.006}{ch.\ 4}
		

\bibitem{Alicki1979}
	R.~Alicki,
	``The quantum open system as a model of the heat engine'',
	J.\ Phys.\ A: Math.\ Gen.\ \href{https://doi.org/10.1088/0305-4470/12/5/007}{{\bf 12} L103}, (1979)
	
\bibitem{Kosloff1984}
	R.~Kosloff,
	``A quantum mechanical open system as a model of a heat engine'',
	J.\ Chem.\ Phys.\ \href{https://doi.org/10.1063/1.446862}{{\bf 80}, 1625} (1984)

\bibitem{wheel}
	D.~Dundas, E.~J.~McEniry, and T.~N.~Todorov,
	``Current-driven atomic waterwheels'',
	Nat.\ Nanotechnol.\ \href{https://doi.org/10.1038/nnano.2008.411}{{\bf 4}, 99} (2009)
	
\bibitem{Oppen}
	N.~Bode, S.~V.~Kusminskiy, R.~Egger, and F.~von Oppen,
	``Scattering theory of current-induced forces in mesoscopic systems'',
	Phys.\ Rev.\ Lett.\ \href{https://doi.org/110.1103/PhysRevLett.107.036804}{{\bf 107}, 036804} (2011)
	[arXiv:1103.4809 [cond-mat.mes-hall]]
	
\bibitem{AL}
	R.~Alicki and K.~Lendi,
	{\it Quantum Dynamical Semigroups and Applications}, 2nd ed.,
	Lect.\ Notes Phys.\ \href{https://doi.org/10.1007/3-540-70861-8}{{\bf 717}, 1} (2007)

\bibitem{Austin}
	R.~T.~McAdory, Jr. and W.~C.~Schieve,
	``On entropy production in a stochastic model of open systems'',
	J.\ Chem.\ Phys.\ \href{https://doi.org/10.1063/1.435120}{{\bf 67}, 1899} (1977)
	
\bibitem{Spohn}
	H.~Spohn,
	``Entropy production for quantum dynamical semigroups'',
	J.\ Math.\ Phys.\ \href{https://doi.org/10.1063/1.523789}{{\bf 19}, 1227} (1978)
	
\bibitem{q-amps}
	D.~Gelbwaser-Klimovsky, R.~Alicki, and G.~Kurizki,
	``Work and energy gain of heat-pumped quantized amplifiers'',
	EPL \href{https://doi.org/10.1209/0295-5075/103/60005}{{\bf 103}, 60005} (2013)
	[arXiv:1306.1472 [quant-ph]]

\bibitem{flywheel}
	A.~Levy, L.~Di\'osi, and R.~Kosloff,
	``Quantum Flywheel'',
	Phys.\ Rev.\ A \href{https://doi.org/10.1103/PhysRevA.93.052119}{{\bf 93}, 052119} (2016)
	[arXiv:1602.04322 [quant-ph]]
	
\bibitem{autorotor}
	A.~Roulet, S.~Nimmrichter, J.~M.~Arrazola, S.~Seah, and V.~Scarani,
	``Autonomous Rotor Heat Engine'',
	Phys.\ Rev.\ E \href{https://doi.org/10.1103/PhysRevE.95.062131}{{\bf 95}, 062131} (2017)
	[arXiv:1609.06011 [quant-ph]]
	
\bibitem{on-off}
	N.~Platt, E.~Spiegel, and C.~Tresser,
	``On-off intermittency: A mechanism for bursting'',
	Phys.\ Rev.\ Lett.\ \href{https://doi.org/10.1103/PhysRevLett.70.279}{{\bf 70}, 279} (1993)
	
\bibitem{Hopf}
	K.~Mallick and P.~Marcq,
	``Stability analysis of a noise-induced Hopf bifurcation'',
	Eur.\ Phys.\ J.\ B \href{https://doi.org/10.1140/epjb/e2003-00324-y}{{\bf 36}, 119} (2003)
	[arXiv:cond-mat/0312360 [cond-mat.stat-mech]]
	
\bibitem{Frisch}
	U.~Frisch,
	{\it Turbulence:  The Legacy of A.~N.~Kolmogorov},
	(Cambridge: Cambridge University Press, 1995)
	
\bibitem{Chandler}
 	A.~Jenkins,
	``Chandler wobble: Stochastic and deterministic dynamics'',
	in {\it Dynamical Systems: Theoretical and Experimental Analysis}, ed.\ J.\ Awrejcewicz,
	Springer Proc.\ Math.\ Stat.\ \href{https://doi.org/10.1007/978-3-319-42408-8_15}{{\bf 182}, pp.\ 177--186} (2016)
	[arXiv:1506.02810 [physics.geo-ph]]
	
\bibitem{macro}
	A.~Jenkins,
	``Towards a microeconomic theory of the finance-driven business cycle'',
	arXiv:1312.0323 [q-fin.GN]
	
	
\bibitem{Groszkowski}
	J.~Groszkowski,
	{\it Frequency of Self-Oscillations},
	(Oxford: Pergamon Press, 1964), sec.\ 9.4
	
\bibitem{Harrison}
	D.~Harrison,
	``Analysis of the Mechanisms for Compensation in Clock B'',
	in {\it Harrison Decoded: Towards a Perfect Pendulum Clock}, eds.\ R.~McEvoy and J.~Betts,
	(Oxford: Oxford University Press, 2020), \href{https://doi.org/10.1093/oso/9780198816812.003.0010}{pp.\ 149--74}
	
\bibitem{uncertainty}
	A.~C.~Barato and U.~Seifert,
	``Thermodynamic Uncertainty Relation for Biomolecular Processes'',
	Phys.\ Rev.\ Lett.\ \href{http://doi.org/10.1103/PhysRevLett.114.158101}{{\bf 114}, 158101} (2015)
	[arXiv:1502.05944 [cond-mat.stat-mech]]

\bibitem{oscillation-cost}
	Y.~Cao, H.~Wang, Q.~Ouyang, and Y.~Tu,
	``The free-energy cost of accurate biochemical oscillations'',
	Nat.\ Phys.\ \href{http://doi.org/10.1038/nphys3412}{{\bf 11}, 772} (2015)
	[arXiv:1506.05686 [physics.bio-ph]]
	
\bibitem{tribo}
	R.~Alicki and A.~Jenkins,
	``Quantum theory of triboelectricity'',
	Phys.\ Rev.\ Lett.\ \href{https://doi.org/10.1103/PhysRevLett.125.186101}{{\bf 125}, 186101} (2020)
	[arXiv:1904.11997 [cond-mat.mes-hall]]
	
\bibitem{Ramaswamy}
	S.~Ramaswamy,
	``The Mechanics and Statistics of Active Matter'',
	Annu.\ Rev.\ Condens.\ Matter Phys.\ \href{https://doi.org/10.1146/annurev-conmatphys-070909-104101}{{\bf 1}, 323} (2010)
	[arXiv:1004.1933 [cond-mat.soft]]
	
\bibitem{Cates}
	T.~Markovich, \'E.~Fodor, E.~Tjhung, and M.~E.~Cates,
	``Thermodynamics of Active Field Theories: Energetic Cost of Coupling to Reservoirs'',
	Phys.\ Rev.\ X \href{https://doi.org/10.1103/PhysRevX.11.021057}{{\bf 11}, 021057} (2021)
	[arXiv:2008.06735 [cond-mat.stat-mech]]
	
\end{thebibliography}
\end{document}